\documentclass[10pt,journal,twoside,shade,web]{ieeecolor}
\usepackage{generic}
\usepackage{amsmath,amssymb,amsfonts}
\usepackage{graphicx}
\usepackage{textcomp}
\usepackage[backend=bibtex,style=numeric]{biblatex}
\usepackage{mathrsfs}
\usepackage{amssymb}
\usepackage{amsfonts}
\usepackage{amsbsy}
\usepackage{latexsym}
\usepackage{amsmath} 
\usepackage{epsfig,psfrag,color}
\usepackage[ruled,linesnumbered]{algorithm2e}
\SetKwRepeat{Do}{do}{while}%
\usepackage{graphicx}

\usepackage[noend]{algpseudocode}
\usepackage{booktabs}
\usepackage{siunitx}
\usepackage{hyperref}
\usepackage{courier}
\usepackage{comment}

\usepackage{caption}
\usepackage{lipsum}
\usepackage{multirow}
\usepackage{diagbox} 
\usepackage{subcaption}
\usepackage{biblatex}
\usepackage{amsthm}

\newcommand{\E}{\mathbb{E}}
\newcommand{\R}{\mathbb{R}}
\newcommand{\tE}{\mathcal{E}}
\newcommand{\x}{x}

\newcommand{\br}{{\rm r}}
\newcommand{\X}{x}

\newcommand{\W}{{w}}
\newcommand{\V}{{\bf V}}
\newcommand{\w}{w}

\newcommand{\Prb}{\mathbb{P}}

\newcommand{\A}{{\bf A}}
\newcommand{\B}{{\bf B}}
\newcommand{\C}{{\bf C}}
\newcommand{\F}{{ F}}

\newcommand{\Hf}{\mathcal{H}}

\newcommand{\Cm}{\mathcal{C}}
\newcommand{\Sigm}{\sigma_\mathcal{M}}
\newcommand{\Dj}{\mathcal{D}_j}
\newcommand{\Cov}{\text{Cov}}
\newcommand{\Var}{\text{Var}}
\newcommand{\Ltw}{{L^{2, *}_{{\rho_T}}}}
\newcommand{\bgp}{\bar g_\psi}

\definecolor{amber}{rgb}{1.0, 0.49, 0.0}
\newcommand{\CR}{\color{red}}

\newcommand{\norm}[1]{\left\lVert#1\right\rVert}
\newcommand{\abs}[1]{\left\lvert#1\right\rvert}

\makeatletter
\newsavebox{\@brx}
\newcommand{\llangle}[1][]{\savebox{\@brx}{\(\m@th{#1\langle}\)}%
  \mathopen{\copy\@brx\kern-0.5\wd\@brx\usebox{\@brx}}}
\newcommand{\rrangle}[1][]{\savebox{\@brx}{\(\m@th{#1\rangle}\)}%
  \mathclose{\copy\@brx\kern-0.5\wd\@brx\usebox{\@brx}}}
\makeatother

\newcommand\vertarrowbox[3][2ex]{%
	\begin{array}[t]{@{}c@{}} #2 \\
		\left\uparrow\vcenter{\hrule height #1}\right.\kern-\nulldelimiterspace\\
		\makebox[0pt]{\scriptsize#3}
	\end{array}%
}

\newtheorem{thm}{\bf Theorem}
\newtheorem{rem}{\bf Remark}
\newtheorem{lem}{\bf Lemma}
\newtheorem{Prop}{\bf Proposition}
\newtheorem{defn}{\bf Definition}

\newtheorem{asm}{\bf Assumption}
\newtheorem{exmp}{\bf Example}

\bgroup

\usepackage[english]{babel}
\usepackage[utf8]{inputenc}
\usepackage[T1]{fontenc}
\usepackage[colorinlistoftodos]{todonotes}
\usepackage{dsfont}
\usepackage{bbm}

\newcommand{\nader}[1]{{\color{blue} #1}}
\newcommand{\arash}[1]{{\color{amber} #1}}

\usepackage[normalem]{ulem}

\def\BibTeX{{\rm B\kern-.05em{\sc i\kern-.025em b}\kern-.08em
    T\kern-.1667em\lower.7ex\hbox{E}\kern-.125emX}}

\begin{document}

\title{Learning Nonlinear Couplings in Network of Agents from a Single Sample Trajectory}

\author{Arash Amini$^{1}$, Qiyu Sun$^{2}$ and Nader Motee$^{1}$
	\thanks{$^{1}$A. Amini and N. Motee are with Department of Mechanical Engineering and  Mechanics, Lehigh University, Bethlehem, PA 18015, USA
		{\tt\small (a.amini,motee)@lehigh.edu}}%
	\thanks{$^{2}$Q.Sun is with the Department of Mathematics, University of Central Florida, Orlando, FL 32816, USA
		{\tt\small qiyu.sun@ucf.edu}}%
}

\maketitle

\begin{abstract}
We consider a class of stochastic dynamical networks whose governing dynamics can be modeled using a coupling function. It is shown that the dynamics of such networks can generate geometrically ergodic trajectories under some reasonable assumptions. We show that a general class of coupling functions can be learned using only one sample trajectory from the network. This is practically plausible as in numerous applications it is desired to run an experiment only once but for a longer period of time, rather than repeating the same experiment multiple times from different initial conditions. Building upon ideas from the concentration inequalities for geometrically ergodic Markov chains, we formulate several results about the convergence of the empirical estimator to the true coupling function. Our theoretical findings are supported by extensive simulation results. 
\end{abstract}


\section{Introduction}
    Interaction among members of a community plays a crucial role in the emergence of holistic behaviors in various natural and engineering systems ranging from interacting atoms to form complex molecules, herd of bison, social networks, the platoon of self-driving cars, interconnected power networks, evolution, and reform mechanisms in financial markets.
    These interactions can be through swarm physics, e.g., flow interaction in the school of fish \cite{keith-2021}, or common control objectives, e.g., potential function-based robot navigation\cite{dan-potential}.
    There have been several fundamental studies to understand and model interactions in some of these systems \cite{vicsek1995novel,cucker2007emergent,cucker2007mathematics,STROGATZ20001,strogatz1994norbert}, where the standard approach is to leverage the underlying logic and physics of such systems and obtain a proper coupling function, by trial and error, in order to replicate collective dynamic behavior of these systems using computer simulations. 
    
    The recent advancements in statistical learning theories { \cite{mohri2018foundations,cucker2007learning}} combined with the dramatic growth of computational power have provided a solid foundation to learn large-scale dynamics from  extensive sensory data in an end-to-end manner. This has lately resulted in intensive research in areas related to learning dynamical systems \cite{poggio2002mathematical,hardt2016gradient,dean2020sample,brunton2016discovering,dean2019safely}. One of the major challenges in learning dynamical networks is that the existing methods suffer from curse of dimensionality, i.e., computation becomes very expensive and the learning accuracy deteriorates rapidly as network size increases. If the underlying dynamics have a certain structure, then one may hope to reduce learning computational and handle large-scale networks. For instance,  in some applications the dynamics of the entire network can be inferred by learning a common coupling function among the agents    \cite{bongini2017inferring,lu2019nonparametric}. The class of networks with gradient-type interaction laws are investigated in \cite{bongini2017inferring}, where it is shown that increasing the number of agents will improve the approximation accuracy. A class of homogeneous and heterogeneous networks is considered in  \cite{lu2019nonparametric,lu2021learning}, where it is shown that  agents' coupling functions can be learned using {\it multiple} sample trajectories and  the learning accuracy improves as the number of trials increases. 
    The closest work in spirit to ours is  \cite{chen2021maximum}, where the authors prove that one can learn the coupling functions with only one trajectory in the presence of Gaussian noise if the interaction law is of gradient type and the dynamics of the resulting network is \textit{linear}. In this paper, we address the problem of learning a {\it general} class of coupling functions in the presence of bounded stochastic noise using only one {\it single} sample trajectory. 
    
    The idea of learning dynamical systems using a single sample trajectory \cite{foster2020learning} is  practically plausible as  in numerous applications it is more feasible and cost effective to collect samples from an ongoing experiment, rather than repeating the same experiment multiple times from different initial conditions. 

Our main distinct contributions with respect to the existing literature are twofold. First, we prove that a class of stochastic dynamical networks can generate geometrically ergodic trajectories under some reasonable assumptions, i.e., the joint evolution probability distribution of trajectories will converge geometrically fast to an invariant stationary distribution; see Theorem \ref{thm.network.ergodic}. This development is necessary to ensure that collecting new sample points along the same trajectory will contain useful information for learning purposes. Building upon this result, in our second main contribution, we prove that for such geometrically ergodic stochastic dynamical networks, one can learn a class of nonlinear coupling functions using only one {\it single} sample trajectory over a convex and compact Hypothesis space that satisfies the coercivity condition; see Theorem \ref{thm.learn.coercive}. It is shown that as the length of the sample trajectory increases, the learning accuracy enhances accordingly up to its limit with high confidence levels. In this case, the approximation error, i.e., the distance between the true coupling function and its approximation, tends to the distance between the true coupling function and the space of hypothesis functions. If the true coupling function lies inside the space of hypothesis functions, then the approximation error will tend to zero; see Theorem \ref{thm.learn.coercive}. Finally, when the class of bounded Lipschitz functions is adopted as the hypothesis space and the true coupling function belongs to this space, we provide an upper bound for the expected value of the learning error and quantify the convergence rate for the error functional in terms of the length of the sample trajectory (see Theorem \ref{thm.convergence.rate}). 


\vspace{0.1cm}
\noindent {\it Mathematical Notations:}
We employ operator $;$ to concatenate two column vectors $x,y \in \R^d$ to obtain column vector $[x;y] \in \R^{2d}$. The symbol $\otimes$ represents the Kronecker product  and \linebreak[4]$\mathbbm{1}=[1,\ldots,1]^T \in \R^n$ is the vector of  all ones. 

\section{Problem Statement}	\label{problemstatement.section}

Let us consider a class of stochastic interconnected dynamical networks whose dynamics are governed  by 
\begin{equation}\label{eqn.network.dyn.agent}
    \x^{i}_{t+1} = \x^{i}_t +h\sum_{j = 1}^{n} k_{ij} \phi \big(\|\x^j_t-\x^i_t\|\big) \big(\x^j_t-\x^i_t\big) + h \w^{i}_t
\end{equation}
    with initial condition $x^i_0$, for all $t\in {\mathbb Z}_+$ and $i \in \{1,\ldots,n\}$, 
    where $h$ is the sampling time, $\x^{i}_t \in \R^d$ is the state of agent $i$ at time instant $ht$, and  $\w^{i}_t\in \R^d$  represents the stochastic effect of environment on the dynamics of agent $i$ at time $ht$, which is assumed to be independent of $x_t^i$ for all $t \geq 0$. 
    The interaction between agents $i$ and $j$ is modeled by a coupling function $\phi: \R \rightarrow \R$ and their coupling strength by coefficients $k_{ij} \geq 0$. Agents $i$ and $j$ are coupled iff $k_{ij} \neq 0$. By defining the vector of state variables \linebreak[4] $\X_t=[\x^{1}_t; \ldots; \x^n_t]$
     and  noise input  
    $\W_t=[\w^1_t; \ldots; \w_t^n]$, 
     the overall dynamics of the network can be rewritten in compact form
    \begin{equation}\label{eqn.network.dyn.Laplace}
       \X_{t+1} =  ({I} -h L_{\X_t}) \X_t + h\W_t, 
    \end{equation}
   where the $(i,j)$'th entry of the state-dependent Laplacian matrix  of the underlying graph of the network $L_{\X_t}$ is defined as
    \begin{equation}\label{eqn.dfn.laplacian}
    (L_{\X_t})_{ij} = \left\{\begin{array}{ll}
       -k_{ij} \phi(r^{ij}_t) I_d  &  \textrm{if}\ j \neq i ,\\
       & \\
       \sum_{k=1}^N k_{ik} \phi(r^{ik}_t) I_d  & \textrm{if}  \ j=i,
    \end{array} \right.
    \end{equation}
where the relative state of agent $j$ with respect to agent $i$ is defined by $\br^{ij}_t=\x^j_t-\x^i_t$ with $r^{ij}_t=\|\br^{ij}_t\|$. Throughout the paper, we assume that the entries of the initial condition $\X_0$ of network  \eqref{eqn.network.dyn.Laplace} are bounded i.i.d. random variable and $\|\X_0\|\le R_0$ holds almost surely for some constant $R_0 >0$. 
Moreover, it is assumed that each entries of { $\W_t$  are bounded i.i.d. random variable with 
    \begin{equation}\label{dfn.noise.propos} \E[\W_t]= { 0},\  
        \E [\W_t\W_t^T] = I \otimes \Sigma, \ {\rm and\ } \ 
        \|\W_t\|\le {\omega}
    \end{equation}
 for all $t\in {\mathbb Z}_+$, holds almost surely for some constant \linebreak[4] ${\omega} >0$, which satisfies $\sigma^2 := n\mathrm{Tr}(\Sigma) \leq \omega^2$.}
 
 We rewrite \eqref{eqn.network.dyn.Laplace} as
    \begin{equation}\label{eqn.network.dyn.Learning}
        \X_{t+1} = \X_t + h\F_{\phi}(\X_t) + h\W_t,
    \end{equation}
     where 
     \begin{equation}\label{def.F_phi}
         \big(\F_{\phi}(\X_t)\big)_i = -\sum_{j = 1}^{n} k_{ij} \phi(r^{ij}_t)\br^{ij}_t.
     \end{equation}
For a given coupling function $\psi:{\mathbb R}\longmapsto {\mathbb R}$, let us define  random variable
    \begin{eqnarray} \label{learn.const.single}
        \tE_{\X_t}(\psi) &:= &\frac{1}{N_e} \norm{\frac{\X_{ t+1}-\X_{t}}{h} - \F_{\psi}(\X_t)}^2 \nonumber\\
       &= & \frac{1}{N_e}
       \norm{\F_{\psi-\phi}-\W_t}^2,
    \end{eqnarray}
where $N_e$ is the number of edges in the underlying graph of the network. The empirical error for a given candidate function \linebreak[4] $\psi: \R \rightarrow \R$ can be formulated as 
    \begin{equation}\label{dfn.empirical_error}
        \tE_{T}(\psi):=\frac{1}{TN_e}\sum_{t=1}^{T}  \norm{\frac{\X_{ t+1}-\X_{t}}{h} - \F_{\psi}(\X_t)}^2,
    \end{equation}
where $T$ stands for the length of the sample  trajectory. Our goal is to learn coupling  function $\phi$ by solving optimization problem
	\begin{equation}\label{Optm.problem}
		\begin{aligned}
\hat\phi_T = \arg~	 \underset{\psi\in {\mathcal H} }{\mathrm{minimize}} ~ \tE_{T}(\psi)
		\end{aligned}.
	\end{equation}


	
	

The {\it problem} is to learn coupling function $\phi:\R \rightarrow \R$ using a single,  long enough,  sample trajectory  $\X_{0}, \X_{1}, \cdots, \X_{T}$ of the dynamical network \eqref{eqn.network.dyn.Laplace} by solving \eqref{Optm.problem} and show that by increasing the length of the sampled trajectory $T$ the distance between $\hat\phi_T$, the optimal solution of  \eqref{Optm.problem}, and the true coupling function $\phi$, known as the approximation error, will tend to the distance of $\phi$ from $\mathcal H$. If  $\phi \in \mathcal H$, then the approximation error will tend to zero; see Theorem \ref{thm.learn.coercive}.


\section{Properties of the Dynamical Network}
In order to guarantee certain properties for a dynamical network \eqref{eqn.network.dyn.agent}, the true coupling function and the underlying graph of the network  should satisfy some conditions.

    \begin{asm}\label{asm.true.kernel}
        The coupling function  $\phi$ is positive, 
        continuous on $[0, \infty)$,    differentiable at zero, and upper-bounded by  $S_0 >0$. 
    \end{asm}

The underlying weighted graph of network \eqref{eqn.network.dyn.Laplace} with Laplacian matrix $L_\X$ is denoted by ${\mathcal G}_\X$.  Since the coupling function $\phi$ is positive, there exists an edge between agents $i$ and $j$ in ${\mathcal G}_\X$  if and only if $ k_{ij}>0$, which does not depend on $\X$. Thus, one only needs to evaluate properties of graph ${\mathcal G}_{0}$, i.e.,  ${\mathcal G}_{x}$ at $x=0$.  

    \begin{asm}\label{asm.graph}
        The graph ${\mathcal G}_0$ 
        is undirected, simple, and connected. 
    \end{asm}

The eigenspace of the Laplacian $L_\X$ associated with the eigenvalue zero is the diagonal subspace 	
	 \begin{equation}\label{def.Delta}
	    \Delta = \big\{  [u; \dots; u]\big| u \in \R^d\big\}.
	\end{equation}
Every vector $\X \in \R^{dn}$ can be decomposed as $x=\bar{x}+x_{\perp}$, { where 
$$\bar x:= \frac{1}{n}\mathbbm{1} \otimes \left(\sum_{i=1}^n \x^i\right),$$  
is its projection onto $\Delta$,
and $\X_\perp$ is its  projection onto $\Delta^{\perp}$, the orthogonal complement of  $\Delta$ in $\R^{dn}$.} We can accordingly decompose the dynamics of network  \eqref{eqn.network.dyn.Laplace} as
	\begin{equation}
	  \label{eqn.network.dyn.perp}
	  \X_{t+1, \perp}=
	     (I -h L_{\X_{t, \perp}}) \X_{t, \perp} + h\W_{t, \perp},
	\end{equation}
	and
	\begin{equation}  \label{eqn.network.dyn.Delta}
        \bar  \X_{t+1} =  \bar \X_{t} + h\bar \W_{t}.
	\end{equation}
The projection of $\w_t$  onto $\Delta^\perp$ is given by
    \begin{equation*}
        \W_{t, \perp} := \W_t - \bar \w_t,
    \end{equation*}
    which has expected values $0$ and 
    \[
    \E \big[\w_{t,\perp}\w_{t,\perp}^T \big]   = M_n \otimes \Sigma,
    \]
     where $M_n={ I}- \frac{1}{n} {\mathbbm{1}} {\mathbbm{1}}^T$ is the centering matrix.
    The dynamical system \eqref{eqn.network.dyn.Delta} is independent of the coupling function $\phi$, which implies that its trajectories do not contain any useful information to learn the coupling function $\phi$. Therefore, only the orthogonal part of the trajectories are informative. For simplicity of our notations, we rewrite \eqref{eqn.network.dyn.perp} as    
	\begin{equation}\label{eqn.network.dyn.perp.main}
	  \X_{t+1} = (I -h L_{\X_{t}}) \X_{t} + h\W_{t}, 
	\end{equation}
    where $\X_t\in \Delta^{\perp}${, and update $\E[\w_{t}\w_{t}] = M_n \otimes \Sigma$}. The eigenvalues of $I -h L_{x_t}$ are contained in  $[1-h\lambda_{\max}(L_{x_{t}}) , 1- h\lambda_2(L_{x_{t}}) ]$, where $\lambda_2(L_{x_{t}})$ and
    $\lambda_{\max}(L_{x_{t}}) $ are smallest non-zero and the largest eigenvalue of $L_{x_t}$, respectively.

	\begin{asm} \label{asm.eigen}
         The dynamics of  \eqref{eqn.network.dyn.perp.main} is uniformly contractive  on $\Delta^\perp$, or  equivalently, the maximum Laplacian eigenvalue satisfies  $ \lambda_{\max}(L_{x_{t}}) < \frac{2}{h}$ along all trajectories of the system.
    \end{asm}

This assumption implies that $h$ should be chosen sufficiently small and that 
\if	 
    {\CR \textbf{  (15) is ambiguous: the right hand side depends on $x_t$. but $\zeta$ does not?} infact zeta is depondent on x but we need that zeta be always smaller than $1$ to be lipshitz coefficient which is counteractive. 
    \[
    0 \le \lambda_2 \le \lambda_{\max}
    \]
    we have three cases 
    \begin{enumerate}
        \item $h\lambda_{\max} < 1$ in this case we have
            \[
            \zeta =1-h\lambda_2
            \]
            since $0<h\lambda_2<1$ then we have 
            \[\zeta <1\]
        \item $h\lambda_{2} > 1$ in this case we have
            \[
            \zeta =h\lambda_{\max}-1 
            \]
            in this case if $\zeta < 1$ then we need to have
            \[ h\lambda_{\max}-1 <1 \]
            i.e.
            \[ \lambda_{\max} <2/h \]
            
        \item $h\lambda_{2} < 1$, and $h\lambda_{\max}>1$  in this case we have
        \[
        \zeta =\max\{1-h\lambda_2,h\lambda_{\max}-1\}<1
        \]
        now we have two cases if $1-h\lambda_2>h\lambda_{\max}-1$
        \[
        \zeta =1-h\lambda_2<1
        \]
        otherwise 
        \[
        \zeta =h\lambda_{\max}-1<1
        \]
        which yield to the same condition as second case above.
    \end{enumerate}
    }
\fi     
all eigenvalues of 
    ${ I}-h L_\X$ are contained in $(0,\, 1)$,
     which means that dynamics  \eqref{eqn.network.dyn.perp.main} is uniformly contractive 
    on $\Delta^\perp$. In fact, the spectral radius of the Laplacian satisfies 
    \begin{eqnarray} 
        \zeta  := \sup_{t \geq 0} \max\left\{{\abs{1-h\lambda_2(L_{x_t})},\abs{1-h\lambda_{\max}(L_{x_t})}}\right\} < 1. \label{xi-value}
    \end{eqnarray}

\begin{exmp}\label{example.networktype}
To better understand the above assumption, one can verify that the dynamics of network \eqref{eqn.dfn.laplacian} will satisfy Assumption \ref{asm.eigen} if $\mathcal{G}_x$ is a complete graph and  $ h<\frac{1}{nKS_0}$ or $\mathcal{G}_x$ is a path graph and $h < \frac{1}{2KS_0}$, where $S_0$ is defined in Assumption \ref{asm.true.kernel} and $K=\max \big\{ k_{ij}~\big | ~1\le i,j\le n\big\}$. In general,  system  \eqref{eqn.network.dyn.perp.main} is uniformly contractive on $\Delta^\perp$ if 
\[ h\leq \frac{1}{\Tilde{K}S_0}, \]
where $\Tilde{K}=\max_{1\leq i \leq n} \sum_{j=1}^n k_{ij}$.



\end{exmp}


\begin{Prop}    \label{lem.state.bounded} 
Suppose that dynamical network        \eqref{eqn.network.dyn.perp.main}, which is considered over $\Delta^\perp$,  satisfies Assumption \ref{asm.eigen}, its initial condition is an i.i.d. random variable that is bounded by $R_0$, and its noise input $\W_t$  satisfies  \eqref{dfn.noise.propos} for every $t \geq 0$. Then, 
        \begin{equation}\label{ineq.distnaces.upperbound}
            r^{ij}_t \le 
            2\left(R_0+  \frac{h\omega}{1-\zeta}\right),
        \end{equation}
for every $1\le i, j\le n$ and $t \geq 0$, holds  almost surely.
	\end{Prop}

The result of this proposition asserts that 
the relative positions of agents in presence of bounded noise will remain bounded. In fact, it can be shown that the distance between every pair of agents is bounded by  
\begin{equation}\label{eqn.dfn.R}
    R:= 2\left(R_0+\frac{h{\omega}}{1-\zeta}\right).
\end{equation}

\begin{figure*}[ht]
        \centering
        \begin{subfigure}[t]{.24\linewidth}
            \centering
        	\includegraphics[width=\linewidth]{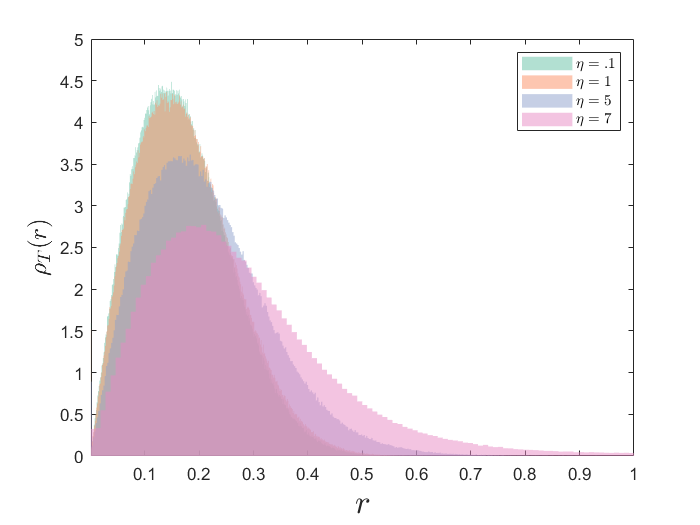}
        	\caption{Coupling Strength}
        	\label{fig:ex1.prb.dist.a}
        \end{subfigure}
        \begin{subfigure}[t]{.24\linewidth}
            \centering
        	\includegraphics[width=\linewidth]{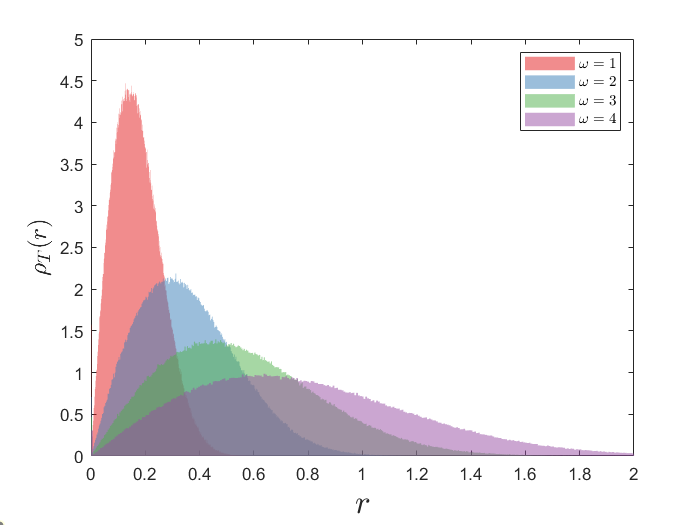}
        	\caption{Noise amplitude}
        	\label{fig:ex1.prb.dist.b}
        \end{subfigure}
        \begin{subfigure}[t]{.24\linewidth}
            \centering
        	\includegraphics[width=\linewidth]{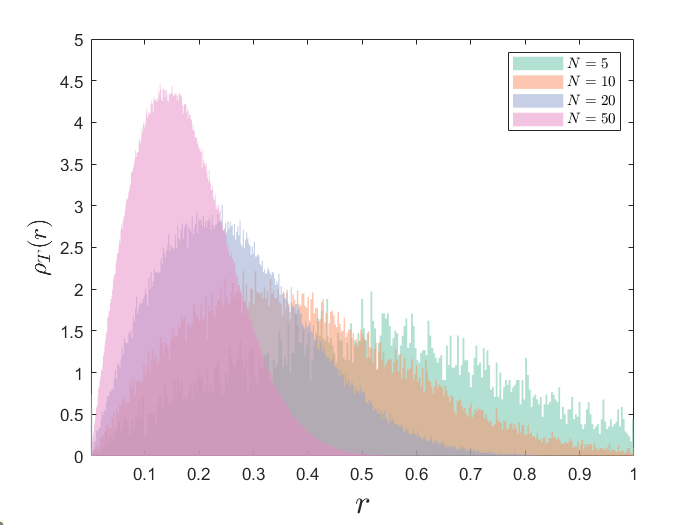}
        	\caption{Number of Agents}
        	\label{fig:ex1.prb.dist.c}
        \end{subfigure}
        \begin{subfigure}[t]{.24\linewidth}
            \centering
        	\includegraphics[width=\linewidth]{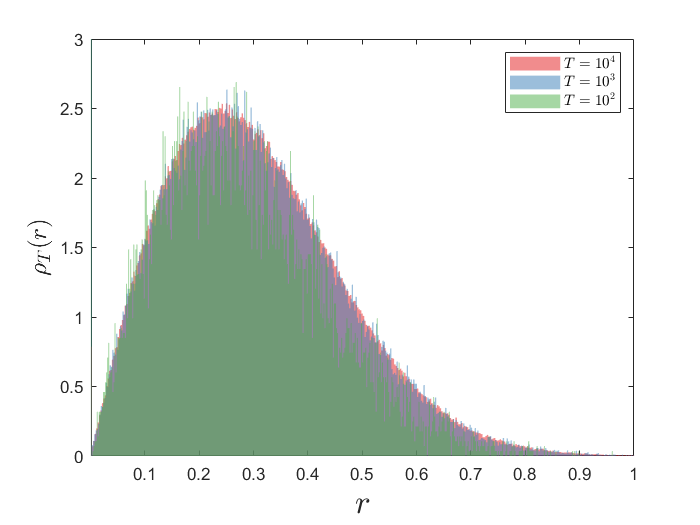}
        	\caption{Sampling Length}
        	\label{fig:ex1.prb.dist.d}
        \end{subfigure}
        \caption{ Empirical probability  distribution, $\rho_T$, for different scenarios discussed in Example \ref{exm.Cucker.Smail} }
        \label{fig.ex1.prb.dist}
    \end{figure*}

\section{Geometric Ergodicity of the  Network}
    
     
Following the problem statement in Section \ref{problemstatement.section}, our goal is to learn the coupling function from a single sample trajectory. In order to   ensure that the empirical estimator converges to the expected estimator as the number of samples are increased, we prove that the Markov chain  \eqref{eqn.network.dyn.Learning}, under some technical assumptions, is  geometrically ergodic \cite{cline1999geometric,tong1990non}. Let us consider the  stochastic  process 
\begin{equation}\label{eqn.Markov.chain}
		x_{t+1} = G(x_t,w_{t})
\end{equation}
that generates a Markov chain $\{x_t\}_{t \geq 0}$ in $\mathcal{X} \subset \R^{nd}$, where $w_t \in \R^{nd}$  are i.i.d random variables with the marginal distribution that is given by a lower semi-continuous density function $\mathfrak{g}$ with respect to a Lebesgue measure that has support $\mathcal{W}=\{\w \in \mathcal{X}| \mathfrak{g}(w) >0\}$. The $t$'th step transition probability of this Markov chain  is defined by 
	\begin{equation} \label{eqn.Markov.transition}
		\Prb^t(x,A)= \Prb(x_t \in A |~ x_0 = x),
	\end{equation}
for $x \in \mathcal{X}$ and  $A \in \mathbb{B}$, in which  $\mathbb{B}$ is the set of all Borel sets \cite{gardiner1985handbook}.
	
	\begin{defn} \label{Definition.ergodicity}
	The stochastic process 
	$\{x_t\}_{t \geq 0}$ generated by  \eqref{eqn.Markov.chain} is geometrically ergodic if there exists a probability measure $\pi$ on $(\R, \mathbb{B})$, a number   $0 < \rho <1 $, and a $\pi$-integrable nonnegative measurable function $h: \R^d \rightarrow \R$ such that
		\begin{equation}\label{def.ergodic}	
			\norm{\Prb^t(x,\cdot)-\pi(\cdot)}_{TV} \leq \rho^t h(x),
		\end{equation}
		where 
		\begin{equation*}
		    \norm{\Prb^t(x,\cdot)-\pi(\cdot)}_{TV} = \sup_{A\in\mathbb{B}} \abs{\Prb^t(x,A)-\pi(A)},
		\end{equation*}
		is the total variation norm  \cite{tong1990non,cline1999geometric}.
	\end{defn}
	
This definition implies \cite{tong1990non} that 
the joint probability distribution of a geometrically ergodic Markov chain converges exponentially to a stationary probability distribution $\pi$.

    \smallskip
	
	\begin{thm}\label{thm.network.ergodic}
Suppose that dynamical network \eqref{eqn.network.dyn.perp.main} satisfies Assumptions  \ref{asm.true.kernel}, \ref{asm.graph}, and \ref{asm.eigen}. Then,	
        the  stochastic process $\{\X_t\}_{t \geq 0}$ in $\Delta^\perp$ generated by \eqref{eqn.network.dyn.perp.main} is geometrically ergodic.
	\end{thm}

	Theorem \ref{thm.network.ergodic} states that the Markov chain evolving according to   \eqref{eqn.network.dyn.perp.main} is geometrically ergodic, which implies that there exists an invariant probability measure $\pi$ independent of the initial condition that satisfies \eqref{def.ergodic}.  We define the empirical probability distributions $\rho_T: [0,\,R] \rightarrow \R_+$ for network \eqref{eqn.network.dyn.perp.main} as 
	\begin{equation}\label{dfn.measure.T}
		    \rho_T(r) := \frac{1}{ N_eT} \sum_{t=0}^{T-1} \sum_{i=1}^{n} \sum_{j\in \mathcal{N}_i} \delta_{r_t^{ij}}(r),
	\end{equation} 
	where ${\mathcal N}_i$ is the set of all neighboring agents of agent $i$ in graph ${\mathcal G}_0$ and    
	\begin{equation*}
	    \delta_{r_t^{ij}}(r) = 
        \begin{cases}
        0 & \text{if } r \neq r_t^{ij}\\ 
        1 & \text{if } r = r_t^{ij}
        \end{cases}.
	\end{equation*}
By applying the law of large numbers (see Theorem \ref{thm.LLN} in the appendix), it follows that 
	\begin{equation}\label{dfn.measure.infty}
		\rho(r) :=\lim_{T\to \infty} \rho_T(r) = \frac{1}{ N_e} \sum_{i=1}^{n} \sum_{j\in \mathcal{N}_i} \E_\pi\big[\delta_{r_t^{ij}}(r)\big].
	\end{equation}
Moreover, 
	\begin{equation}\label{Cn.condition}
	    \lim_{T\to \infty}
	\frac{1}{T} \sum_{t=0}^{T-1} \|F_\psi(\X_{t})\|^2=
	\E_\pi \big[ \|F_\psi(\X)\|^2 \big],
	\end{equation}
	where   $\pi$ is the invariant probability measure from Theorem \ref{thm.network.ergodic}.

For a given $r \in [0,R]$, the quantity $\rho(r)$ measures those fraction of neighbors whose relative distances are equal to $r$. Thus, for a given interval $J \subset [0,\,R]$, one can utilize $\rho(J)$ to measure how spread our samples are over $[0,R]$: the larger the value of $\rho(J)$, the more distinct (informative) the samples.

\begin{exmp}\label{exm.Cucker.Smail}
Let us consider the swarm dynamics  \eqref{eqn.network.dyn.agent} with $d=2$, in which   $\mathcal{G}_0$ is a complete graph and the coupling function is 
\begin{equation}\label{exm.cucker.smail.function}
    \phi(r)=\frac{\Gamma}{(1+r^2)^\eta}.
\end{equation}
This is widely known as the Cucker-Smale coupling function \cite{cucker2007emergent}. Figure \eqref{fig.ex1.prb.dist} illustrates the effect of noise amplitude $\omega$, length of sampling $T$, coupling strength $\eta$, and the number of agents $N$ on the empirical probability distribution $\rho_T$. Each subplot in Figure \eqref{fig.ex1.prb.dist} depicts the empirical measure of the probability distribution $\rho_T$ by setting $\Gamma= 0.4$, $T=10^4$, $\eta =1$, $\omega = 1$,  and $n =50$ as the fixed parameters. 
As it is shown in Figure \ref{fig:ex1.prb.dist.a}, by increasing $\eta$ the coupling strength weakens, which results in a network of agents whose relative distances are more scattered. When the noise amplitude is increased,  inequality \eqref{ineq.distnaces.upperbound} implies that the relative distances may experience larger fluctuations. Figure \ref{fig:ex1.prb.dist.b} shows how the probability distribution $\rho_T$ starts to flatten and spread along the axis as the noise amplitude increases. According to Theorem \ref{thm.ergodic}, if the network is ergodic,  then $\rho_T$ converges to $\rho$ as $T \rightarrow \infty$, which is shown in Figure \ref{fig:ex1.prb.dist.d}. 
\end{exmp} 
In most applications, the number of agents and the coupling function is fixed. Therefore, the length of the sample trajectory controls the accuracy of the probability distribution $\rho_T$, which is illustrated in Example  \ref{exm.Cucker.Smail}. However to modify the range of learning, $R$, one can change noise amplitude, $\omega$, to obtain the desired interval domain for the estimated coupling function.



	\section{Convergence of Learning} \label{learningtheory.section}
	\if
    \nader{This is a very big section. Can we break it into a few Sections or at least put them into a few Subsections ?}	
    \arash{I beilve we can break it into two subsection
    1. Compact hypothesis space
    2. Convex and Coervite hypothesis space}
    \nader{Please check and replace all "kernel" by "coupling function".}
	\fi

The geometric ergodicity property of the network is necessary to define the steady-state empirical probability distribution  $\rho: [0,R] \rightarrow \R_+$, which in turn allows us to define a Hilbert space. Let us consider the space of functions $L^{2,*}_{\rho_T}([0,R])$ as the weighted Hilbert space of  measurable functions with respect to $\rho_T: [0,R] \rightarrow \R_+$ that is endowed by  
    \begin{equation}\label{dfn.l2nrm}
        \|\psi\|_{L^{2, *}_{{\rho_T}}([0,R])}:=
    \left(\int_0^R |\psi (r) r|^2 \rho_T(dr)\right)^{1/2},
    \end{equation}
 for all $T\ge 1$. For clarity, we simply use notation  $L^{2,*}_{\rho_T}$. Using \eqref{def.F_phi}, one gets 
	\begin{equation}\label{ineq.Error.Measure}
	   \E_\pi[\|F_\psi(\X)\|]^2 \leq N_e K^2 \|\psi\|_{L^{2, *}_{{\rho}}}^2,
	\end{equation}
	where $K=\max \big\{ k_{ij}~\big | ~1\le i,j\le N \big\}$.
\if	
	\arash{This inequality provide crucial information regarding learnabilty of this problem. If the above inequality was equality instead, then one can follow similar approach to \cite{cucker2007mathematics} and shows that the compactness and convexity are sufficient assumptions over the hypothesis space $\Hf$ to prove the existence and uniqueness of the estimator. However in our case one need to add another criteria, so-called coercivity condition(see Assumption \ref{asm.coercivity}) to guarantee such properties. }
\fi	
The geometric ergodicity property of the network enables us to apply the law of large numbers to the empirical error functional and show that its expectation exists, which is defined by
    \begin{align}\label{dfn.error.infty}
        \tE(\psi)   &:=   \lim_{T \to \infty} \frac{1}{T}
        \sum_{t=1}^T \tE_{\X_{t}}(\psi)  \nonumber \\
        & =
       \frac{1}{N_e} \left(\E_\pi \left[\norm{\F_{\psi-\phi}(\X)}^2\right] + \sigma^2 \right),
    \end{align}
and is  well-defined for all $\psi\in {\mathcal H}$. Suppose that $\hat \phi$ is the estimator of the expected error functional over the hypothesis space $\Hf$, i.e.,
	\begin{equation}\label{dfn.estimator.infty}
	    \hat{\phi}:= \arg \underset{\psi\in \Hf}{\mathrm{minimize}}~ \tE(\psi).
    \end{equation}
We investigate the convergence of the empirical  estimator $\hat{\phi}_T$ to the true coupling function $\phi$. When $\phi\in {\mathcal H}$, it can be shown that  $\hat \phi=\phi$. For a given parameter $\delta \in (0,1)$,  our goal  is to quantify the minimum trajectory length $T$ to ensure that 
\[\|\hat \phi_T-\hat\phi\|_\Ltw \leq \epsilon,\]
holds with probability $1-\delta$.  First, in Subsection \ref{learningtheory.subsection.compact}, it is shown that if the hypothesis space is compact, the error of the empirical estimator converges to the error of the expected estimator. Then, in Subsection \eqref{learningtheory.subsection.B}, it is proven that if in addition to compactness the hypothesis space satisfies the convexity and coercivity conditions, then the expected estimator is also unique.  

    

    \begin{asm}\label{asm.H.bounded}
        The hypothesis space ${\mathcal H}$ is a bounded subset of $ L^\infty([0,R])$, i.e., 
        \begin{equation}\label{def.SHR}
            S_{{\mathcal H}}:= \sup_{\psi\in {\mathcal H}} \|\psi\|_{L^{\infty}([0,R])}<\infty,
        \end{equation}
        and $S_{\Hf}\geq S_0$. 
    \end{asm}
    All functions $\psi \in \Hf$ has domain $[0,\,R]$, therefor to improve the tractability of our theoretical results we simply use the notation $\norm{\psi}_\infty$  instead of $\|\psi\|_{L^{\infty}([0,R])}$.
    
\begin{rem}
Using Assumption \ref{asm.H.bounded} and \eqref{dfn.l2nrm}, one can show that 
        \begin{equation}\label{ineq.l2.linfty.bounded}
                 \|\psi\|_{L^{2, *}_{\rho_T}}\le  R \|\psi\|_{\infty}<\infty,
        \end{equation}
for every $\psi\in {\mathcal H}$, which reveals a relationship between the  two norms. 
    \end{rem}

    \subsection{Error  Convergence}\label{learningtheory.subsection.compact}
    
 Suppose that the hypothesis  space $\Hf$ is a compact subset of $L^\infty([0,R])$ and the expected estimator lies in $\Hf$. It is shown that using these minimal assumptions one can only prove that { $\tE(\hat{\phi}_T)$} will converge to  { $\tE(\hat{\phi})$} with the desired confidence if the sampled trajectory is long enough. For two given candidate functions $\psi_1,\psi_2 \in \Hf$, we show that the difference of the corresponding empirical errors of $\psi_1$ and $\psi_2$ is always bounded by the $\Ltw$ distance of the two functions.
    
    \begin{lem}\label{lem.upbound}
        For every $\psi_1, \psi_2\in {\mathcal H}$,  
        \begin{align}\label{lem.upbnd.T}
           \abs{\tE_T(\psi_1)-\tE_T(\psi_2)} &\le  2 K^2RS  \| \psi_1-\psi_2\|_{L^{2, *}_{\rho_T}},\\
           \label{lem.upbnd.infty}
           \abs{\tE(\psi_1)-\tE(\psi_2)} &\le  2 K^2RS  \| \psi_1-\psi_2\|_{L^{2, *}_{\rho}}
        \end{align}
        hold almost surly,  where
        $S := S_{{\mathcal H}, R}+\|\phi\|_{\infty} +{\bar{\omega}}/{R}.$
	\end{lem}

To state our next result, we should introduce a few new notations. For every $T\geq 1$ and $\psi \in \Hf$, let us define  functional 
        $$L_T(\psi) := \tE(\psi) - \tE_T(\psi).$$
Then, for every $\psi_1,\psi_2 \in \Hf$, it follows that
\begin{align}\label{ineq.diff.infty}
\abs{L_T(\psi_1)-L_T(\psi_2)} & \leq  4K^2R^2S  \| \psi_1-\psi_2\|_\infty \nonumber \\
& \leq 8K^2R^2S^2,
\end{align}
holds almost surely, where \eqref{ineq.diff.infty} is a direct consequence of\linebreak[4]  Lemma \ref{lem.upbound}. Using \eqref{learn.const.single}, we define  function $g_\psi: \R^{dn} \rightarrow \R$ by 
            \begin{equation}\label{dfn.g.psi}
                g_\psi(\X_t) := \tE_{\X_t}(\psi)-\tE(\psi),
            \end{equation}
for all $\psi \in \Hf$, whose asymptotic variance  is given by
            \begin{equation}\label{dfn.g.asymvar.psi}
                \Sigm^2(\psi) = \Var_\pi {g_\psi(\X_0)} +2\sum_{i=1}^\infty \Cov_\pi \big[g_\psi(\X_0),g_\psi(\X_i)\big].
            \end{equation}
In the next theorem, we prove that when the hypothesis space $\Hf$ is compact and the expected estimator $\hat \phi$ lies in it, then the empirical error converges to the expected error as length of the sampled trajectory, i.e., $T$, tends to infinity.
	
\begin{thm}\label{thm.compact}
Suppose that   $\Hf$ is a compact subset of ${L}^\infty([0,R])$ and   $\hat \phi \in \Hf $. Then, for every $\epsilon > 0$, we have  
	 \begin{align}\label{thm.compact.ineq.prb}
       \hspace{-.5cm}     \Prb\left\{ \abs{\tE(\hat \phi_T) - \tE(\hat \phi)}\leq \epsilon\right\} &\geq \\
            & \hspace{-3cm} 1- C \;  \mathcal{N}\left(\Hf,\frac{\epsilon }{16K^2R^2S}\right)  e^{\left(\frac{-\epsilon^2T}{512\sigma_\Hf^2+32 \tau \epsilon 8K^2R^2S^2 \log(T)}\right)} \nonumber
	    \end{align}
        where $C$ and $\tau$ are constants with respect to ergodidcty rate of \eqref{eqn.network.dyn.perp.main},  $\mathcal{N}(\Hf,\,l)$ is the minimum number of balls with diameter $l>0$ required to cover $\mathcal{H}$ with respect to the {$L^{\infty}$ norm}, and  $\sigma_\Hf^2$ is the supremum of the asymptotic variances of random variables $g_\psi(\X_t)$ , i.e., 
        \begin{equation*}
            \sigma_\Hf^2=\sup_{\psi \in \Hf} \Sigm^2(\psi).
        \end{equation*}
    \end{thm}

	
	\subsection{Convergence of Empirical Estimator}\label{learningtheory.subsection.B}
	 
    	
Suppose that  $\phi \in \Hf$. Then, 
	\begin{equation*}
	    \E [\norm{F_{\psi-\phi}}^2] \leq N_e K^2 \norm{\psi-\phi}_\Ltw^2
	\end{equation*}
for every $\psi \in \Hf$, which is a consequence of \eqref{ineq.Error.Measure}. 
	If the above inequality changes to equality the convexity of the hypothesis space is sufficient to prove the existence of a unique minimizer for \eqref{Optm.problem} \cite{cucker2007learning}. In our case,  however, the convexity of $\Hf$ will not be sufficient and one has to further assume other properties to guarantee that the expected estimator is unique and lies in $\Hf$. The next assumption will enable us to ensure the learnability of our problem.
	
	\begin{asm}  \label{asm.coercivity}
        The hypothesis space ${\mathcal H}$ is a compact and convex set of functions on $\R_+$ that satisfy the  coercivity condition 
         \begin{equation}\label{dfn.coercivity.condition-1}
            c_{\mathcal H} := \frac{1}{N_e} \inf_{\psi \in \Hf \setminus \{0\}} \left\{\frac{ \E_{\pi}  \left[ \norm{\F_\psi(\X)}^2 \right]}{ \norm{\psi}^2_{L^{2, *}_{\rho}}}\right\} >0.
        \end{equation}
\end{asm}

If  $\psi-\phi \in \Hf$, then inequality 
    \begin{equation}\label{rem.coercivity.eq1}
        c_\Hf \norm{\psi-\phi}_{L^{2, *}_{\rho}}^2 \leq \tE(\psi)
    \end{equation}
is a direct consequence of Assumption \ref{asm.coercivity}. Furthermore, one can show that  
    \begin{equation}\label{rem.coercivity.eq2}
        c_\Hf \|\psi-\hat \phi\|_{L^{2, *}_{\rho}}^2 \leq \tE(\psi)-\tE({\hat \phi}).
    \end{equation}
Thus, $\hat \phi$ is the unique minimizer of  $\tE$ over $\Hf$.  

    \begin{thm}\label{thm.learn.coercive}
 Suppose that $\Hf$ is a compact, convex, and coercive subset of the ${ L}^\infty([0,R])$. Then,  the error bound  
        \begin{equation}\label{them.learn.coercive.Prb}
            \|\hat{\phi}_T - {\phi}\|^2_{L^{2, *}_{\rho}} \leq 4 \left(1+\frac{K}{c_{\mathcal H}}\right) \inf_{\psi \in \Hf} \norm{\psi- \phi }^2_{L^{2, *}_{\rho}} + \frac{2\epsilon}{c_{\mathcal H}}
        \end{equation}
        holds with probability at least $1-\delta$, for all $\epsilon > 0 $ and $\delta\in (0, 1)$, provided that the length of the sampled trajectory $T$  satisfies
        \begin{align}\label{thm.learn.coercve.delta.ineq}
        \hspace{-1cm}    C \; \mathcal{N}\left(\Hf,\frac{ \epsilon}{24 K^2 R^2 S}\right) & \nonumber \\
            & \hspace{-3cm}\le \delta \exp\left\{ \frac{T\epsilon}{96K^2S^2R^2 \left(\frac{128  c_\mathcal{M}K^2}{c_\Hf}  + \tau    \log(T)\right) }\right\}, 
        \end{align}
        where $\tau,\, C,\, c_\mathcal{M}$ are positive constant with respect to ergodicity rate of the network \eqref{eqn.network.dyn.perp.main}.
    \end{thm}

The result of Theorem \ref{thm.learn.coercive} asserts that one can learn a coupling function with an arbitrary precision with probability $1-\delta$  if the trajectory length $T$ is long enough.

     \begin{rem}
         If the true coupling function $\phi \in \Hf$, then for all $T \ge 1$, the error bound
         \begin{equation*}
            \|\hat{\phi}_T-{\phi}\|^2_{L^{2, *}_{\rho}} 
	        \leq  \frac{2\epsilon }{c_\Hf}
         \end{equation*}
         holds with probability at least $1-\delta$, where $\delta$ satisfies \eqref{thm.learn.coercve.delta.ineq}.  The  convergence rate of the empirical estimator to the true coupling function can be  controlled by coercivity constant $c_\Hf$.
     \end{rem}

     
     \begin{rem}
         The bias term in \eqref{them.learn.coercive.Prb} solely depends on the choice of the hypothesis space, which emphasizes that the coupling functions can be learned as far as the hypothesis spaces allow.
     \end{rem}
     

    Let us define the set of functions  $\mathcal{K}_{R,S} \in {\bf L}^\infty([0,R])$ by
    \begin{equation}\label{dfn}
        \mathcal{K}_{R,S}:=\left\{\psi \in \mathrm{Lip}_c([0,R])~\big|~   \norm{\psi}_\infty + \text{Lip}(\psi) \leq S\right\},
    \end{equation}
    where $\mathrm{Lip}_c([0,R])$ is the class of Lipschitz  functions with compact support over  $[0,R]$ and ${\rm Lip}(\phi)$ is the Lipschitz constant of  $\psi$ over interval $[0,R]$.
    We finish this section by approximating rate of convergence of the empirical estimator when  $\Hf=\mathcal{K}_{R,S}$, and $\phi \in \Hf$.
    \begin{thm} \label{thm.convergence.rate}
    Suppose that $\hat \phi_T$ is the minimizer of the empirical error functional \eqref{dfn.empirical_error} over hypothesis space $\Hf = \mathcal{K}_{R,S}$ and $\phi \in \mathcal{K}_{R,S}$. Then, there exists a  $0\leq \gamma$ such that 
    \begin{equation}\label{thm.convergence.rate.ineq}
        \E_\pi [\|\hat{\phi}_T - {\phi}\|_{L^{2, *}_{\rho}}] \leq \gamma \sqrt[4]{\frac{128  c_\mathcal{M}K^2 + \tau c_\Hf   \log(T)}{T c_\Hf^2}},
    \end{equation}
    where $\gamma$ is a function of $R,\,S,\,K$.
    \end{thm}

	  Theorem \ref{thm.convergence.rate} shows the convergence rate of the empirical estimator  to the true coupling function using one single sample trajectory depends on both coercivity constant and the covariance of the noise. Intuitively, if there is no noise  or the Markov chain is not geometrically ergodic, constructing a probability distribution is impossible, which results in  learning divergence. 
	  When the model is deterministic or does not enjoy the geometric ergodicity property,  the asymptotic variance is unbounded according to Lemma \ref{lem.variance.bound}. From \eqref{thm.convergence.rate.ineq}, we can observe that no matter how large the length of the trajectory is,  the empirical estimator will not converge.

	 \begin{rem}
	 For long enough trajectories with   \linebreak[4] $T \gg \exp(128 c_\mathcal{M} K^2)$, we have 
	 \begin{equation}
	      \E \left[\|\hat{\phi}_T - {\phi}\|_{L^{2, *}_{\rho}}\right] \leq \bar \gamma \left(\frac{    \log(T)}{T c_\Hf}\right)^{\frac{1}{4}}.
	 \end{equation}
In comparison to the results of learning coupling functions using multiple trajectories \cite{lu2019nonparametric}, we observe that using only one sample trajectory requires more information in order to reach the same accuracy for the desired confidence level.
	 \end{rem}
 
\section{Learning Procedure}
\subsection{Learning Algorithm}
We discuss an algorithm to learn the empirical estimator. Suppose that  $\{\psi_i^\Hf\}_{1\leq q \leq Q} \subset L^\infty([0,R])$ is a prespecified frame elements \cite{christensen2003introduction}. We form the hypothesis space by 
    \begin{equation}\label{sim.dfn.basis}
        \Hf = \left\{ \psi ~\Bigg|~ \psi=\sum_{q=1}^{Q}  \varrho_{q} \psi_q^\Hf ~\textrm{for some}~ \varrho_{1}, \ldots,\varrho_{Q} \in \R  \right\}.
    \end{equation}
For every $\psi \in \Hf$, one obtains
     \begin{equation}\label{sim.eqn.F}
         \{\F_{\psi}(\X_t)\}_i = -\sum_{j = 1}^{n}\sum_{q=1}^{Q}k_{ij}  \varrho_{q} \psi_q^\Hf (r^{ij}_t) \br^{ij}_t.
     \end{equation}
Let us define $v_t = \frac{\X_{t+1}-\X_{t}}{h} $. Then, using  \eqref{dfn.empirical_error}  results in
    \begin{align*}
      \tE_{T}({\psi}) &= \frac{1}{TN_e}\sum_{t=1}^{T}  \norm{v_t - \F_{\psi}(\X_t)}^2\\
         & \hspace{-1cm}= \frac{1}{TN_e}\sum_{t=1}^{T} \sum_{i=1}^n \left\|v_t^i - \left[-\sum_{j = 1}^{n}\sum_{q=1}^{Q}k_{ij}  \varrho_{q} \psi_q^\Hf (r^{ij}_t) \br^{ij}_t\right]\right\|^2.
    \end{align*}
This  problem \eqref{Optm.problem} can be cast as a least-squares problem
    \begin{equation}\label{sim.optim.problem}
    \underset{{\varrho \in \R^Q} }{\mathrm{minimize}} ~ \frac{1}{TN_e}\norm{A_T{ \varrho}-b_T}^2,
    \end{equation}
    where    $ \varrho :=[\varrho_1, \ldots, \varrho_Q]^T$,   $b_T:=[v_1;v_2;\cdots;v_T] \in \R^{ndT}$,
    \[
        A_T= \big[\bar A_1^T, \bar A_2^T, \ldots, \bar A_T^T \big]^T  \in \R^{ndT \times Q},
    \]
 and $\bar A_t \in \R^{nd \times Q}$ is given by
    \[
    \{\bar A_{t}\}_{i,q}= -\sum_{j = 1}^{n} k_{ij} \psi_q^\Hf (r^{ij}_t) \br^{ij}_t
    \]
    for all $ 1\leq q\leq Q$ and $ 1\leq i\leq n$.
    One can write the optimal solution of  \eqref{sim.optim.problem} in closed-from 
    \begin{align}\label{sim.optim.solution}
        \hat {{\varrho}}_T = (A_T^T A_T)^{-1}A_T^T b_T, 
    \end{align}
    which gives us the empirical estimator  
    \begin{equation}
        \hat \phi_T(r) = \sum_{q=1}^{Q}  \hat \varrho_{T,q} \psi_q^\Hf(r). \label{estimator-T}
    \end{equation}

  	\begin{figure*}[ht]
        \begin{subfigure}[t]{.24\linewidth}
            \centering
        	\includegraphics[width=\linewidth]{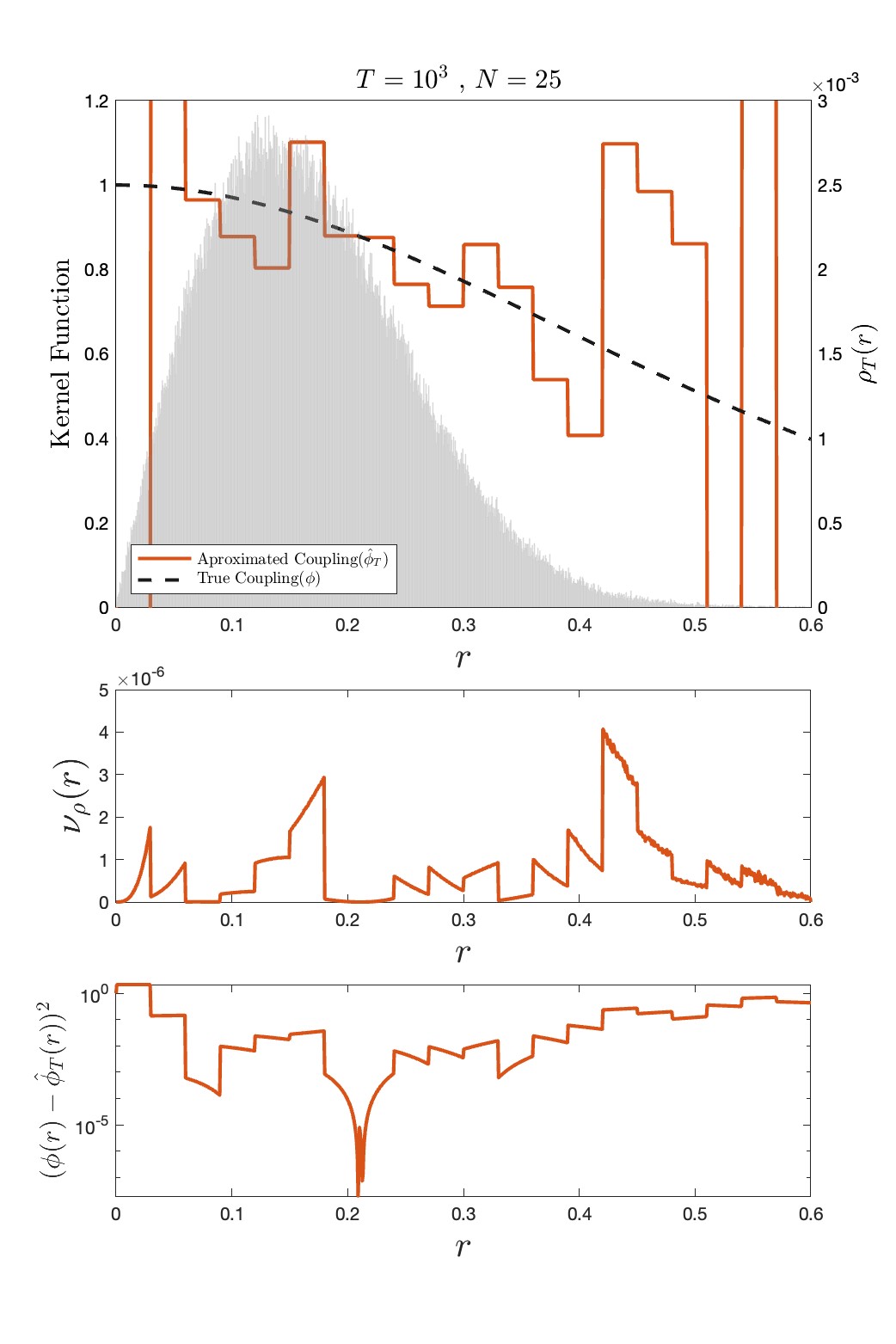}
        \end{subfigure}
        \begin{subfigure}[t]{.24\linewidth}
            \centering
        	\includegraphics[width=\linewidth]{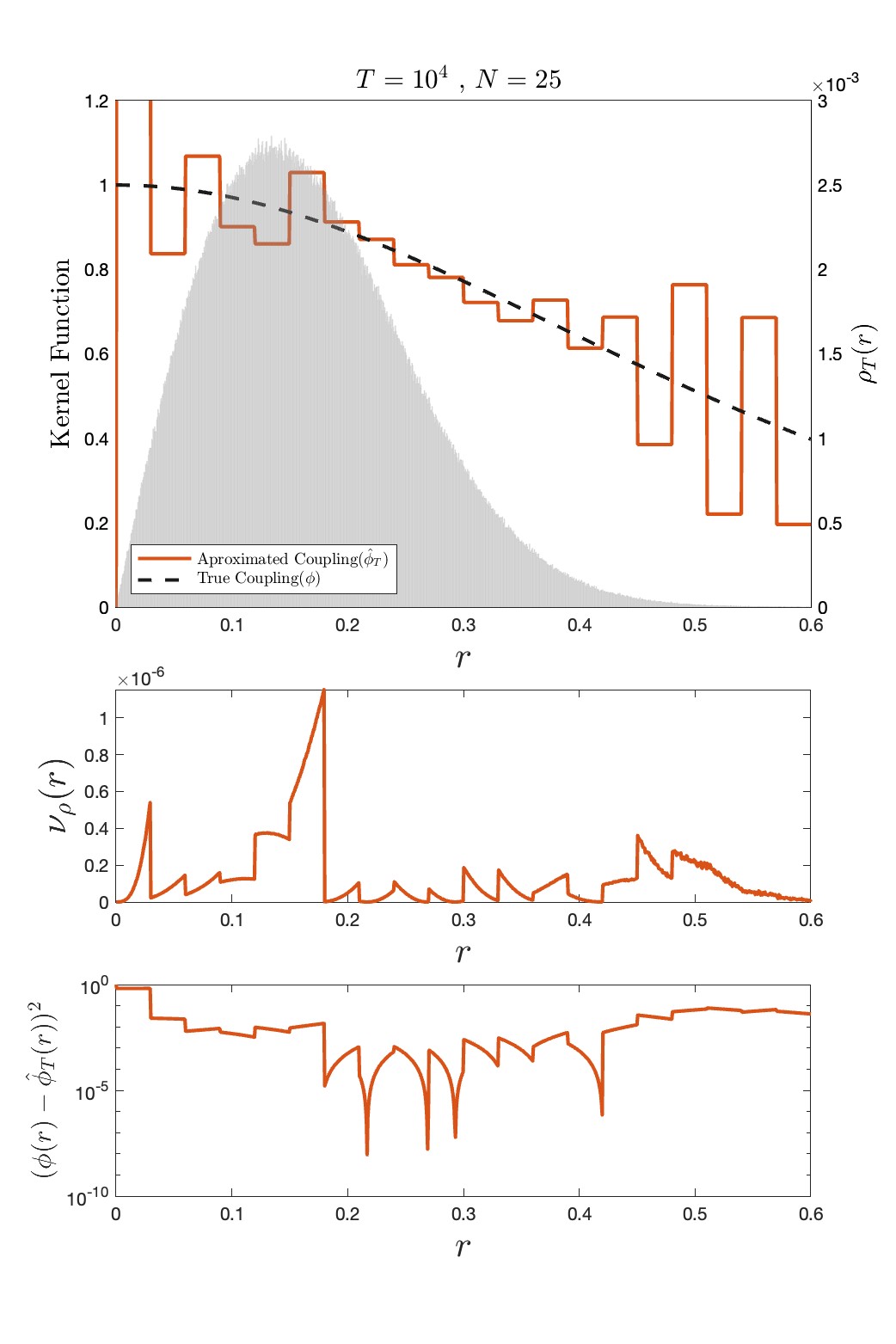}
        \end{subfigure}
        \begin{subfigure}[t]{.24\linewidth}
            \centering
        	\includegraphics[width=\linewidth]{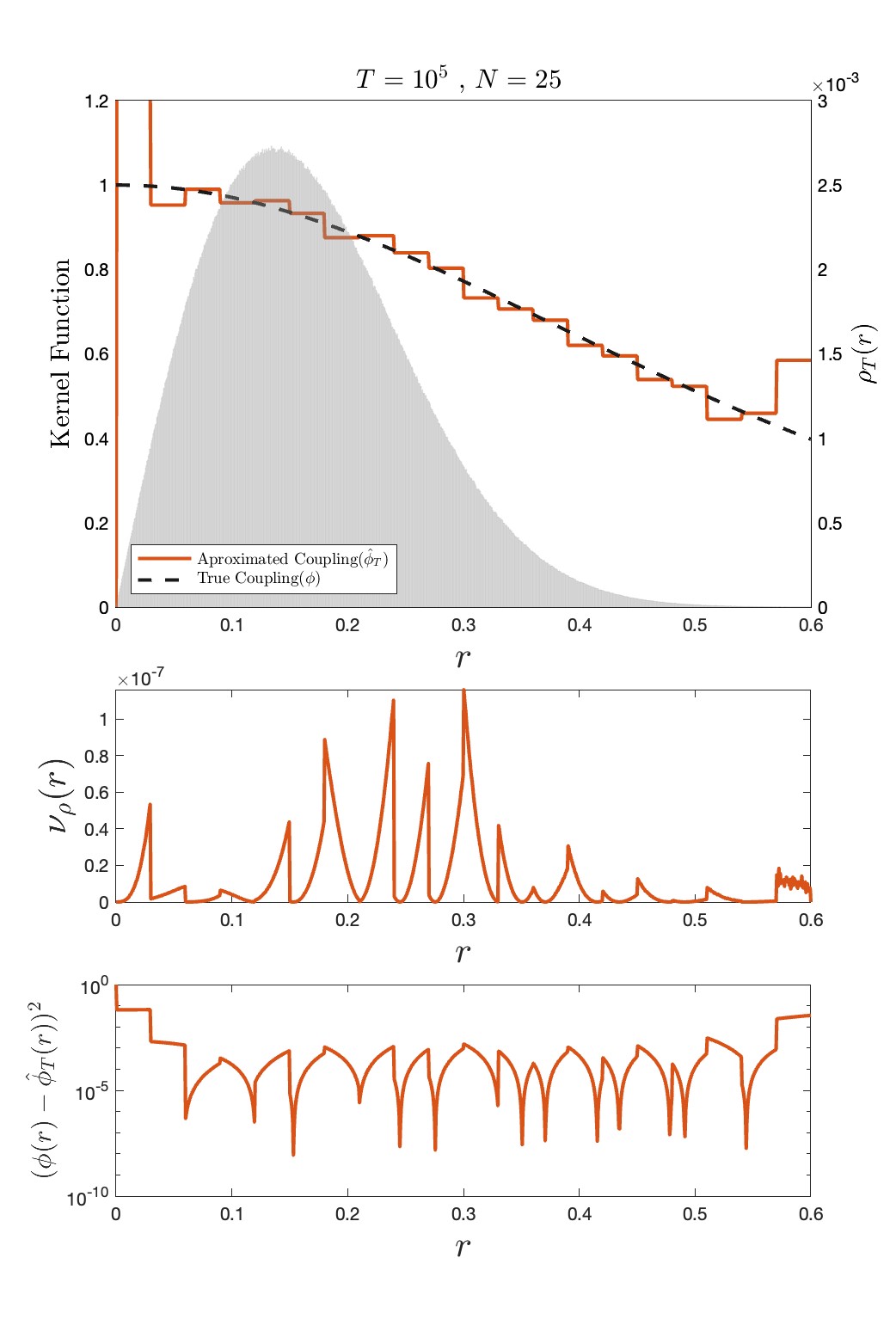}
        \end{subfigure}
        \begin{subfigure}[t]{.24\linewidth}
            \centering
        	\includegraphics[width=\linewidth]{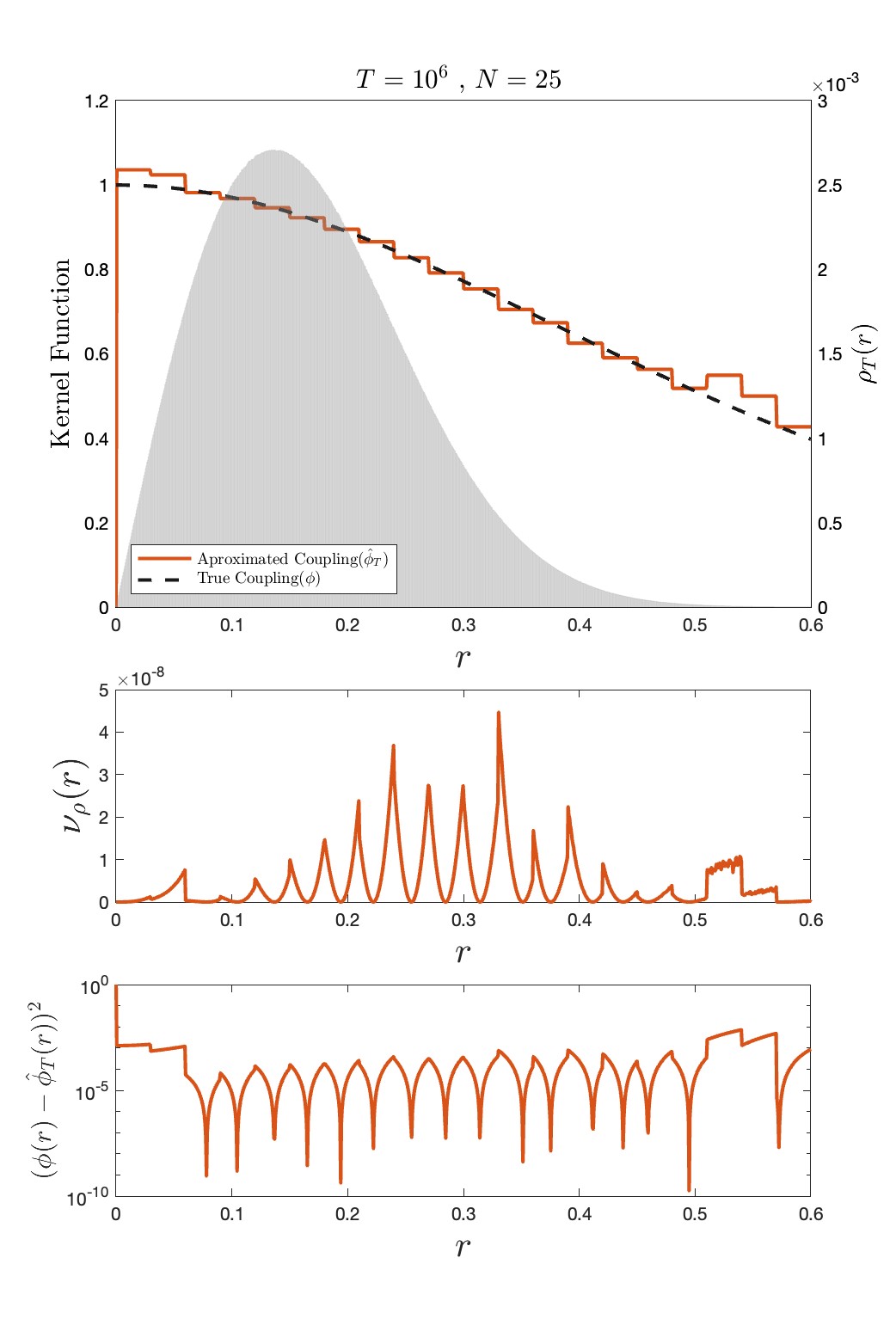}
        \end{subfigure}
                \caption{
                This figure illustrate the \textit{true} coupling function $\phi$, \textit{approximated} couplings $\hat{\phi}_T$ \& \textit{initial} function guess $\phi_0$ and  \textit{probability density} function $\rho_T$, the \textit{point-wise} distance of estimated function and original kernel, $\nu_{\rho,\hat \phi_T}(r)$ and the squared error, $(\phi-\hat\phi_T)^2$,   with different trajectory length $T$ for the first example in Subsection \ref{sim.example.A}}
        \label{fig.ex1.apx}
    \end{figure*}

\subsection{Explicit Form of Coercivity Constant}\label{subsec.coercivity}


We obtain an explicit form for the coercivity condition when the hypothesis space $\Hf$ is  \eqref{sim.dfn.basis}. When there is no prior knowledge about the coupling function, one needs to derive the probability density function $\rho$ empirically. We recall that the coercivity constant is defined by \eqref{dfn.coercivity.condition-1}. 
Knowing that a candidate coupling function can be represented as $\psi=\sum_{q=1}^{Q}  \varrho_{q} \psi_q^\Hf \Phi_0$, it follows that  
 \begin{align*}
     \E_\pi \left [ \left \|\,F_\psi(x)\right\|^2\right ] &= \E_\pi \left [\sum_{i=1}^n ~ \left\|\sum_{j=1}^n k_{ij}\psi(r^{ij})\br^{ij} \right\|^2\right]\\
     &=\E_\pi \left [\sum_{i=1}^n \left\|\sum_{j=1}^n \sum_{q=1}^Q k_{ij}\varrho_q \psi_q^\Hf(r^{ij})\br^{ij} \right\|^2\right] \\
     &=\E_\pi \left [\sum_{i=1}^n \varrho^T \Upsilon^\Hf_{i} \varrho   \right] = \varrho^T \Upsilon_\Hf  \varrho,
 \end{align*}
 where $\Upsilon_\Hf \in \R^{Q\times Q}$ is  a positive semidefinite matrix that is defined by
\[\Upsilon_\Hf = \E_\pi \left [\sum_{i=1}^n  \Upsilon^\Hf_{i} \right]\]
and the elements of $\Upsilon^\Hf_{i} \in \R^{Q\times Q}$ are 
 \begin{align*}
     &\left(\Upsilon^\Hf_i\right)_{qq'} = 
     \left(\sum_{j=1}^n  k_{ij} \psi_q^\Hf(r^{ij})\br^{ij} \right)^T
      \left(\sum_{j'=1}^n  k_{ij'} 
     \psi_{q'}^\Hf(r^{ij'})\br^{ij'} \right).
 \end{align*}
Similarly, one can evaluate the denominator term in \eqref{dfn.coercivity.condition-1} and obtain
 \begin{align*}
\|\psi\|^2_{L^{2, *}_{\rho}}&= \int_0^R (\psi (r) r)^2 \rho(dr)\\
&= \int_0^R \left (\sum_{q=1}^Q \varrho_q \psi_q^\Hf(r) r\right )^2 \rho(dr) = \varrho ^T \Xi_\Hf^T \varrho,
 \end{align*}
in which $\Xi_\Hf^T \in \R^{Q \times Q}$ is  given by 
\[
\left(\Xi_\Hf\right)_{qq'} = \int_0^R  \psi_{q'}^\Hf(r)\psi_q^\Hf(r) r^2 \rho(dr)
\]
for all $q,q' \in \{1,\,2,\, \cdots,\,Q\}$. By definition,   $\Xi_\Hf$ is a positive semidefinite matrix.  Assume that $\varrho_*$ is an eigenvector of  $\Xi_\Hf$ that corresponds to a zero eigenvalue. From \eqref{dfn.l2nrm}, one has
\begin{equation}\label{eqn.orthogonal.subspace}
    \varrho_*^T \Xi_\Hf \varrho_* = \int_0^R \left(\psi_*(r) r\right)^2 \rho(dr) = 0,
\end{equation}
where $\psi_*=\sum_{q=1}^Q \varrho_{*,q} \psi_q^\Hf$. Equation \eqref{eqn.orthogonal.subspace} shows that $\psi_*$ is either zero or orthogonal to the probability measure $\rho$. In other words, the subset of the hypothesis space $\Hf$ that are orthogonal to the $\rho$, which are not informative for learning purposes, are the null space of $\Xi_\Hf$, hence we shall exclude them from learning. From the above derivation, one can conclude that 
\begin{align}\label{dfn.coercivity.Matrix}
    c_\Hf = \inf_{\varrho \notin \ker({\Xi_\Hf}) } \frac{\varrho^T \Upsilon_\Hf \varrho}{\varrho^T \Xi_\Hf \varrho},
\end{align}
where $\ker({\Xi_\Hf})$ is the null-space of $\Xi_\Hf$, which is minimizing the generalized Rayleigh quotient. Suppose that $\Xi_\Hf^{\frac{1}{2}}$ is  the Cholesky decomposition of $\Xi_\Hf$, i.e., $\Xi_\Hf= \Xi_\Hf^{\frac{1}{2}}{\Xi_\Hf^{\frac{1}{2}}}^T$. Then, the coercivity constant can be obtained through
\begin{equation}\label{eqn.coerciv.const.final}
    c_\Hf = \lambda_{\min} \left((\Xi_\Hf^{\frac{1}{2}} )^{-1}\Upsilon_\Hf(\Xi_\Hf^{\frac{1}{2}})^{-T}\right).
\end{equation}
The hypothesis space $\Hf$ will not be coercive if  $\Upsilon_\Hf$ has zero eigenvalues where their corresponding eigenvectors do not belong to $\ker(\Xi_\Hf)$.

The computation of matrices $\Upsilon_\Hf$ and $\Xi_\Hf$ requires a complete knowledge of the probability distribution $\rho$. Since the agent distances are known from the samples, one can empirically approximate  $\Upsilon_\Hf$ and $\Xi_\Hf$, even though the probability distribution $\rho$ is unknown.

	\begin{figure*}[ht]
        \begin{subfigure}[t]{.24\linewidth}
            \centering
    	\includegraphics[width=\linewidth]{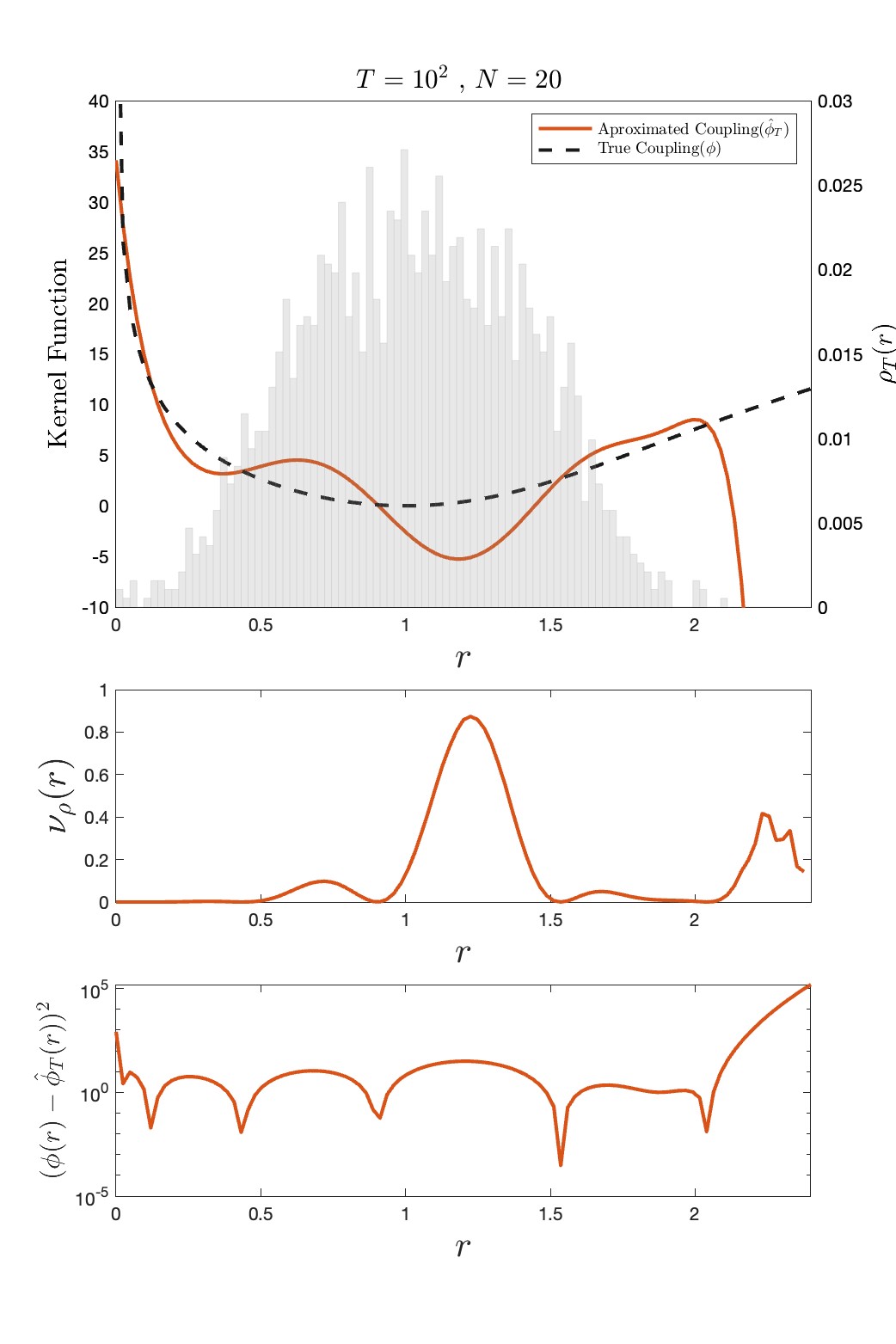}
        	\label{fig.Eg2.T0}
        \end{subfigure}
        \begin{subfigure}[t]{.24\linewidth}
            \centering
    	\includegraphics[width=\linewidth]{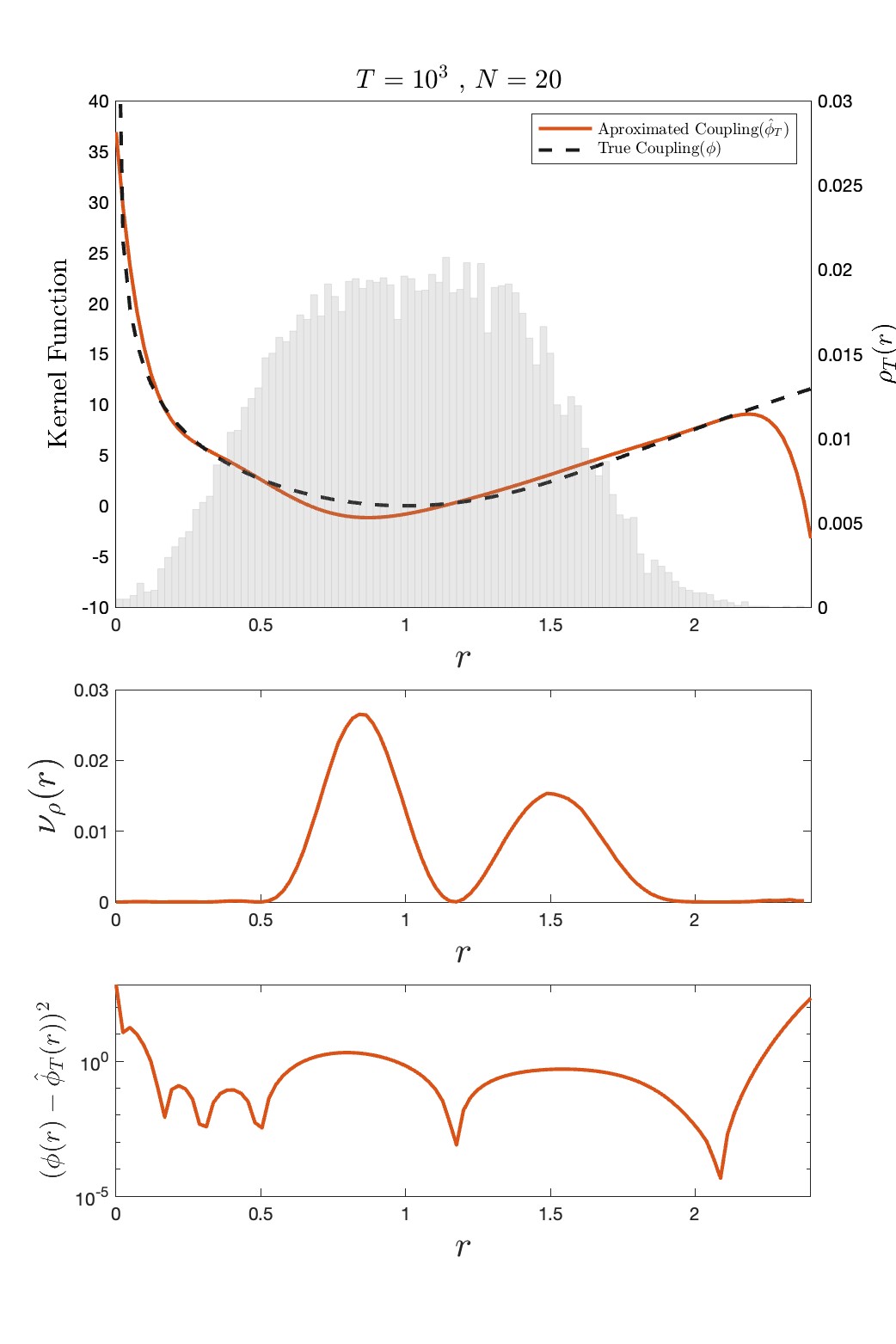}
        	\label{fig.Eg2.T1}
        \end{subfigure}
        \begin{subfigure}[t]{.24\linewidth}
            \centering
    	\includegraphics[width=\linewidth]{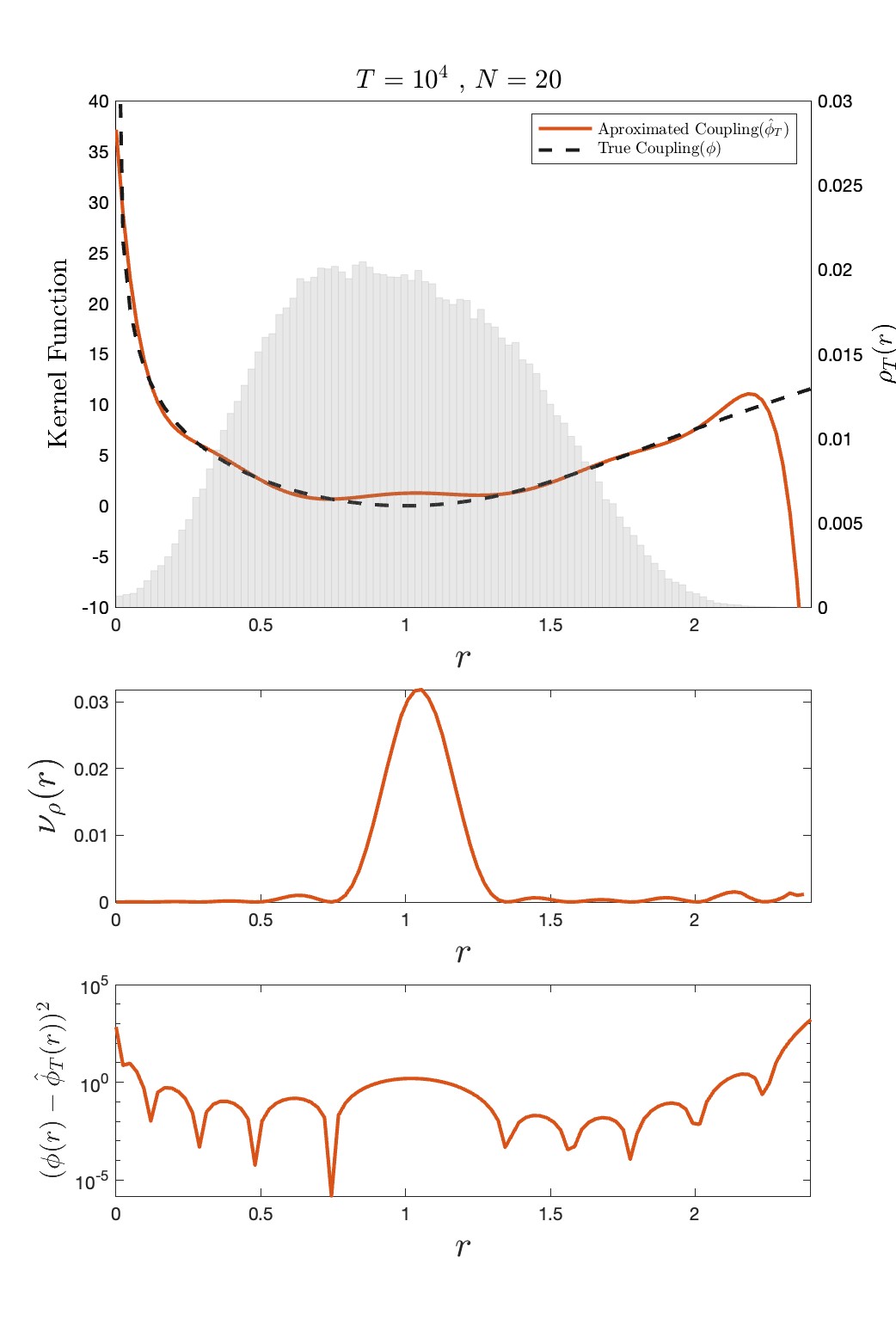}
        	\label{fig.Eg2.T2}
        \end{subfigure}
        \begin{subfigure}[t]{.24\linewidth}
            \centering
    	\includegraphics[width=\linewidth]{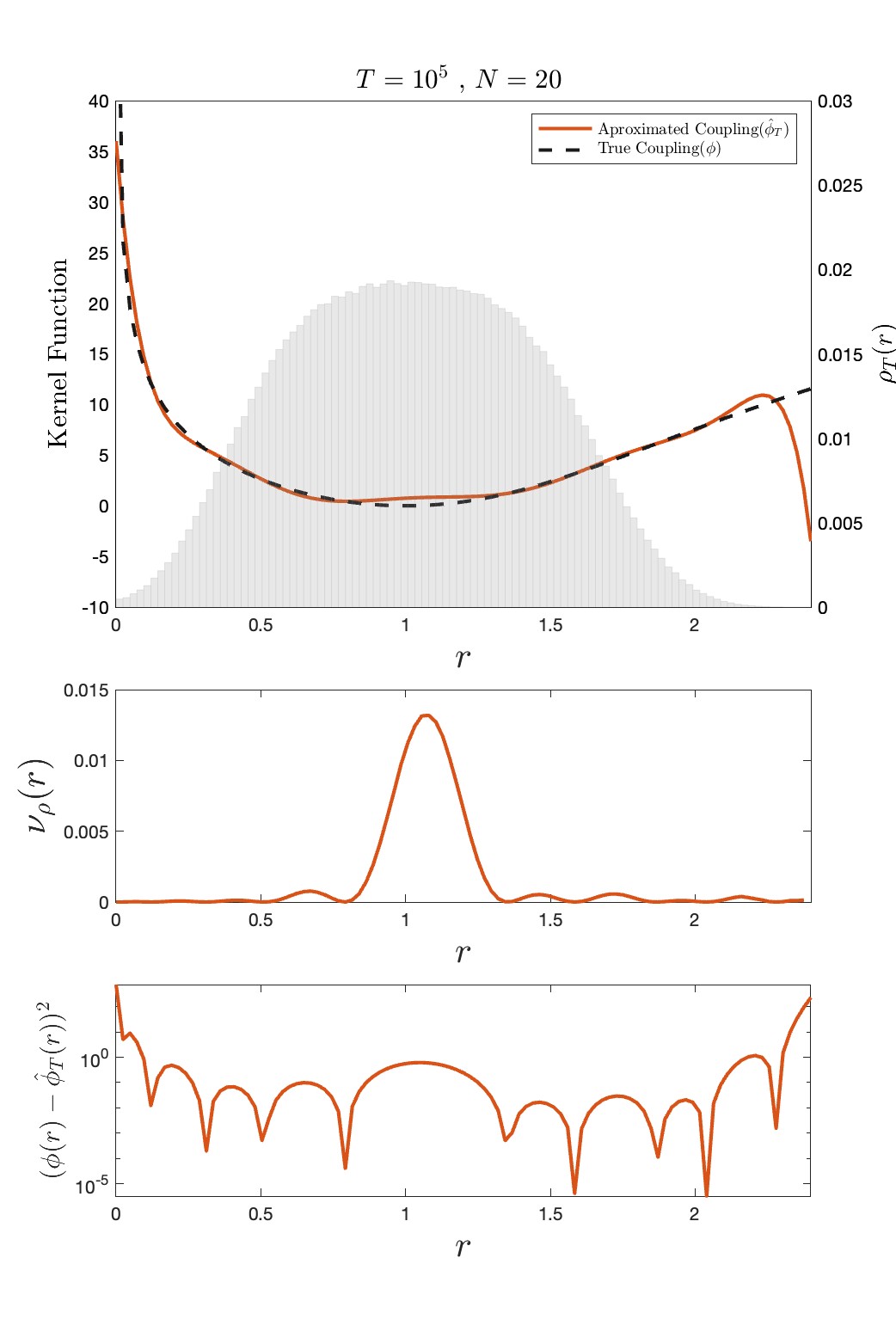}
        	\label{fig.Eg2.T3}
        \end{subfigure}
        \caption{This figure illustrate the \textit{true} coupling function $\phi$, \textit{approximated} couplings $\hat{\phi}_T$ \& \textit{initial} function guess $\phi_0$ and  \textit{probability density} function $\rho_T$, the \textit{point-wise} distance of estimated function and original kernel, $\nu_{\rho,\hat \phi_T}(r)$ and the squared error $(\phi-\hat\phi_T)^2$,   with different trajectory length $T$ for the case study in Subsection \ref{sim.example.b}}
        \label{fig.Eg2}
    \end{figure*}	

	\begin{figure}[t]
        \centering
        \includegraphics[width=\linewidth]{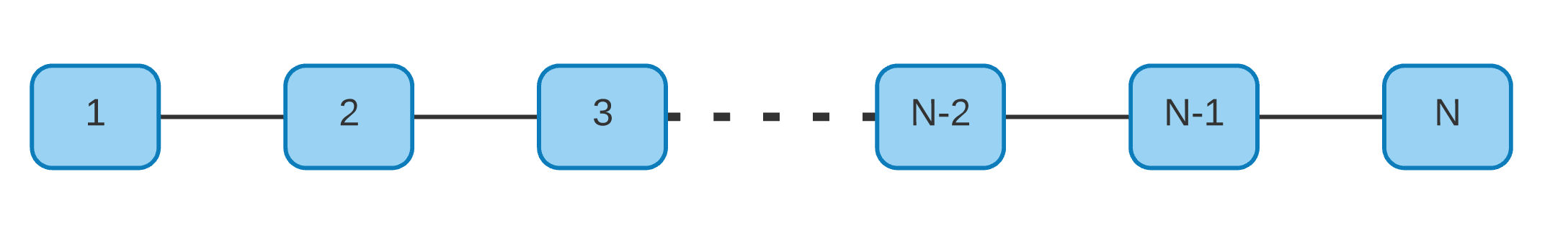}
        \caption{The network's interconnection topology in Subsection \ref{sim.example.b}. }
        \label{fig.Eg2.Network}
    \end{figure}

\section{Simulation Results}
To validate our theoretical results, we provide numerical studies for two different classes of dynamical networks, namely, the first-order version of the  Cucker-Smale's consensus network model \cite{cucker2007emergent} and a first-order dynamical network inspired by the formation control problem, in which agents start from a random initial condition and attempt to maintain a predetermined distance from their nearby neighbors.

Metric \eqref{dfn.l2nrm} measures the distance between a given function and the original coupling function. This measure  requires integration  over the interval  $[0,\, R]$, which makes the point-wise comparison between the two functions implausible. The square of the pointwise differences of the two functions, i.e., { $(\psi(r)- \phi(r))^2$}, does not consider the fact that convergence is subject to the probability distribution $\rho$. To remedy this issue, we define 
\begin{equation}\label{dfn.nu.rho}
    \nu_{\rho,\psi}(r) \, = \, |\big(\phi(r)-\psi(r)\big)r|^2\rho(r),
\end{equation}
which measures the distance between the candidate coupling function and the original coupling function at every  $r \in [0,\,R]$  
{weighted by} probability density $\rho(r)$. 
In several  applications, e.g., risk analysis and prediction, it is crucial to find a coupling function that can recreate the stochastic characteristics of the original system. To this end, we define the probability distribution $\hat{\rho}_T$ by replacing $\phi$ in \eqref{eqn.network.dyn.agent} with $\hat{\phi}_T$, i.e., the stationary probability distribution of the distances between the agents in swarm dynamics 
    \begin{equation*}
    	\hat \x^{i}_{t+1} = \hat \x^{i}_t +h\sum_{j = 1}^{n} k_{ij} \hat \phi_T \big(\|\hat \x^j_t-\hat \x^i_t\|\big) \big(\hat \x^j_t-\hat\x^i_t\big) + h\w_t^i.
    \end{equation*}
To compare the two probabilities, we employ the Kullback–Leibler divergence measure
    \begin{equation*}
        D_{KL}(\rho\|\hat{\rho}_T) = \int_0^R\rho(r)\log\left( \frac{\rho}{\hat{\rho}_T} \right)
    \end{equation*}
    to quantify how far these two probability distributions are from each other.

	\subsection{Cucker-Smale model}\label{sim.example.A}

    In Example \ref{exm.Cucker.Smail}, we discussed this class of first-order dynamical networks, where they can be characterized  using the following spatially decaying coupling functions \cite{cucker2008flocking}
	\begin{equation}\label{exp.kernel}
	    \phi(r)=\frac{\Gamma}{\big(1+r^2\big)^\eta}\,,
	\end{equation} 
	where parameters $\Gamma,\eta >0$ determine coupling strength between the agents. The state space of each agent is   $\R^2$ and each agent is driven by a uniform bounded noise with mean zero and $\E[\w_i^T\w_i] = (\omega^2/3) I$.  It is assumed that there is no prior knowledge about the coupling function. Thus, the class of simple functions is utilized as the hypothesis space, i.e., every function $\psi \in \Hf$ can be represented as
    \begin{equation*}
        \psi(r) = \sum_{q=1}^Q \varrho_q {\bf 1}_{q}(r),
    \end{equation*}
where the indicator function ${\bf 1}_{q}:\R \rightarrow \{0,1\}$ is defined by 
	\begin{equation*}
	   {\bf 1}_{q}(r) = 
        \begin{cases}
        1 & \text{if } r \in [R\frac{q-1}{Q},\,R\frac{q}{Q})\\
        0 & \text{if }   r \notin [R\frac{q-1}{Q},\,R\frac{q}{Q})
        \end{cases},
	\end{equation*}
    $R = 0.6 $  given by \eqref{eqn.dfn.R}. It is assumed that the initial condition of every agent is zero and that they are all-to-all connected with network parameters $\Gamma=1, \eta= 0.4$, $Q=20$, and noise amplitude $\omega= 10$. We assume that the network is symmetric and therefore all communications have the same weights, i.e, $k_{ij} = 1$, for all $i\neq j$.
    From our discussion in Example \ref{example.networktype} and the fact that graph $\mathcal{G}$ is complete, it follows that if $h<\frac{1}{S_0 n}$,     then the network dynamics \eqref{eqn.network.dyn.agent} is ergodic.
     For the guaranteed ergodicity of network with $25$ agents, the above inequality transforms to $h\, < \,0.04$. Therefore we chose $h=0.01$ to make sure that the sampling time $h$ is sufficiently small to ensure the required ergodicity property of the dynamical network.

	\begin{figure}[t]
        \centering
        	\includegraphics[width=\linewidth]{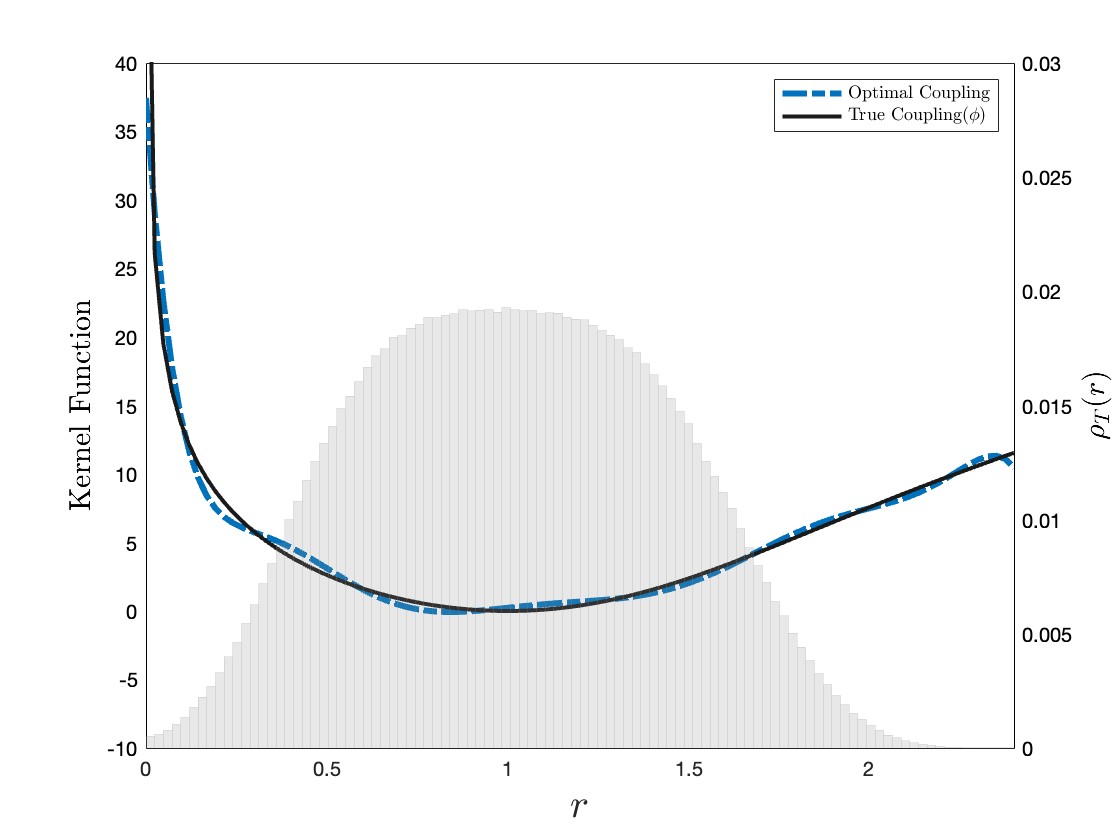}
        \caption{ This figure depicted the
        \textit{original} coupling function $\phi$, 
        probability \textit{density} $\rho$ and the \textit{best approximation} of original couplings, $\phi$, over hypothesis space $\Hf$ for the second example in  Subsection \ref{sim.example.b}}
        \label{fig.ex2.Best}
    \end{figure}

    Figure \ref{fig.ex1.apx}
    shows the empirical estimator $\hat{\phi}_T$, the empirical probability distribution $\rho_T$, the pointwise distance between $\hat{\phi}_T$ and $\phi$ weighted by ${{\rho}}$, i.e.,  $\nu_{\rho,\hat{\phi}_T(r)}$, and the square of the difference between the estimated and the original functions, i.e., $(\hat{\phi}_T-\phi)^2$, for  different sample trajectory lengths $T=\{10^3,\,10^4,\,10^5,\,10^6 \}$.      The empirical estimation of the coupling function i.e., $\hat{\phi}_T$, converges pointwise to the original function as illustrated in Figure \ref{fig.ex1.apx} in the regions where the probability measure $\rho$ is non-zero. 
    The learning algorithm is anticipated to converge more accurately at the probability distribution's peaks as these points are  where the agent's distance information is found in abundance.

     The hypothesis space $\Hf$ is constructed by the indicator functions that are orthogonal to each other, i.e.,
    \[\int_0^R \psi_q(r)\psi_{q'}(r)r^2\rho(dr)=0, \]
    for all $q\neq q'$. Therefore,  $\Xi_M$ is a diagonal matrix. We can further simplify \eqref{eqn.coerciv.const.final} and get
    \[c_\Hf = \lambda_{\min}(\Xi_\Hf^{-1}\Upsilon_\Hf) = 3.8469 > 0,\]
    which indicates that the candidate hypothesis space is coercive and convex. Thus, Theorem \ref{thm.learn.coercive} can be applied, which  asserts that the empirical estimator converges to the original function with respect to  $L^{2, *}_{{\rho}}$--norm, i.e., the convergence depends on the probability density $\rho$ and it is not pointwise. From Figure \ref{fig.ex1.apx}, one can conclude that  $(\hat{\phi}_T-\phi)^2$ can assume relatively large values compared  to $\nu_{\rho,\hat{\phi}_T(r)}$, while their peaks are in different regions.

Table \ref{table:ex1} displays the empirical error $\tE_T(\hat{\phi}_T)$, the distance of the estimated and the original couplings $\|\phi-\hat{\phi}_T\|_{L^{2, *}_{{\rho}}}$, and the  Kullback–Leibler divergence $D_{KL}(\rho\| \hat{\rho})$ for various  trajectory lengths. We remark that the $D_{KL}(\rho\|\hat{\rho}_T)$ is computed empirically from the sample trajectory.
The results in Table \ref{table:ex1} are in line with Theorem \ref{thm.learn.coercive}, where both the  empirical error and the distance between the estimator and the original function are expected to decrease to certain saturation values as $T$ increases.

It should be emphasized that the original coupling function does not belong to the hypothesis space. As a result, the error function $\tE_T(\hat{\phi}_T)$ is expected to reach a saturation level. On the other hand, the Kullback–Leibler 
divergence $D_{KL}(\rho\| \hat\rho_T)$ decreases at a slower rate than the empirical error. This hints that even with small sample trajectories, which result in an estimator that is usually far from the original function, the probability distribution $\hat \rho_T$ can be close to $\rho$ as the network  structure remains intact.

 \begin{table}[t]
        \centering
        \begin{tabular}{||c |c c c||} 
         \hline
         T & $\tE_T(\hat{\phi}_T)$ &  $\|\phi-\hat{\phi}_T\|_{L^{2, *}_{{\rho}}}$ &$D_{KL}(\rho||\hat{\rho}_T)$\\ [1ex] 
         \hline\hline
         $10^2$ & $0.0809$ & $0.0917$ & $0.0013$ \\ 
         $10^3$ & $0.0256$ & $0.0298$ & $7.54\times 10^{-5}$ \\
         $10^4$ & $0.0081$ & $0.0117$ & $5.63\times 10^{-5}$ \\
         $10^5$ & $0.0026$& $0.0073$ & $4.46\times10^{-5} 
         $ \\ 
         $10^6$ & $0.0008$& $0.0021$ & $2.97\times10^{-5}$ \\
         $\infty$ & $0.0004$& $0.0018$ & $2.47\times10^{-5}$ \\[.5ex] 
         \hline
        \end{tabular}
        \caption{This Table illustrates the error functional, $\tE(\hat\phi_T)$, the distance of empirical estimator and the original function, $\|\phi-\hat{\phi}\|_{L^{2, *}_{{\rho_T}}}$ and the Kullback–Leibler divergence, $D_{KL}(\rho||\hat{\rho})$,  
         with different trajectory length $T$ for the first example in Subsection \ref{sim.example.A}}.
        \label{table:ex1}
    \end{table}

	\subsection{Collision Avoidance Formation}\label{sim.example.b}

We consider formation control of a group of agents along the horizontal axis that are coupled through a graph with chain topology  \cite{liu2021risk}. This model appears in several  applications including placement of mobile sensors and surveillance with flying machines such as satellite. To avoid collision between every two consecutive agent, we use a nonlinear coupling that generates repulsive force when the agents are too close to each other and attractive force when they are far away. The control objective is maintain these agents in a certain distance $r_0$ from each other in presence of exogenous noise. 
The underlying communication graph $\mathcal{G}_0$ is an undirected  chain graph , as shown in Fig.  \ref{fig.Eg2.Network}, in which the agents are only allowed to communicate with their nearest neighbors. The dynamics of the network is given by \eqref{eqn.network.dyn.agent} with $d=1$, 
    \begin{equation*}
        k_{ij} = 
        \begin{cases}
            1 & {\rm if} \quad j\in \{i+1,\, i-1\} \\ 
            0 & {\rm if} \quad{\rm otherwise}
        \end{cases},
    \end{equation*}
and coupling function
    \begin{equation}\label{eqn.Ex2.Coupling}
        \phi(r)= \frac{\Gamma (r-r_0)^2}{(a-(r-r_0)^3)^\eta} ,
    \end{equation}
    where $a,\Gamma$, and $\eta$ are positive constants. Figure \ref{fig.ex2.Best} depicts the coupling function and it is assumed that each agent is disturbed by a uniform bounded noise with mean zero and $\E[\w_i^T\w_i] = (\omega^2/3) I$. In order to achieve the desired formation, we change the equilibrium of the swarm dynamics \eqref{eqn.network.dyn.agent} to ${\bf b} = [0,\,r_0,\,2r_0, \cdots,n r_0]$.  It is assumed that there is no prior knowledge about the coupling function and that the hypothesis space is the set of all polynomials with order less than or equal to $Q=10$, i.e., every function $\psi \in \Hf$ can be represented by
    \begin{equation}
        \psi(r)=\sum_{q=1}^Q \varrho_q r^q.
    \end{equation}
Adopting polynomials as our hypothesis space will  allow us to apply the  method introduced in Subsection \ref{subsec.coercivity} to estimate the coercivity condition. After calculations, the coercivity is $c_\Hf = 0.0371$, which shows that the candidate hypothesis space is coercive with respect to probability distribution $\rho$. 
In this case study, the number of agents is $n=20$, the coupling parameters are 
\[\Gamma =10,\quad \eta =0.4,\quad r_0=1,\quad a=1.01, \]
and the initial condition is drawn randomly from the uniform distribution over $[-20,20]$.
It is further assumed that the maximum communication range between the agents is $3$ units
, i.e., $R_0=3$, as a result, the parameter $S_0 \,=\, 43.0957 $. 
    Example \ref{example.networktype} states that in order for  the current path networks to have counteractive dynamics the time step  $h$ needs to satisfy $h \leq 0.0118$, therefore we chose $h\,=\,0.01$. As a consequence of Theorem \ref{thm.network.ergodic} the dynamical network with the proposed coupling function is geometrically ergodic.  The numerical results for different trajectory length $T=10^2,\,10^3,\,10^4,\,10^5$ are depicted in Fig. \ref{fig.Eg2}. The top row shows the probability distribution $\rho_T$, along with the original coupling function $\phi$, and the learned coupling function $\phi_T$. The second row shows the distance between the original  and the learned coupling function with respect to the weighted pointwise  error  \eqref{dfn.nu.rho}. The last row illustrates the  pointwise squared error between the original and the learned coupling functions. For better illustration, the last two rows have different scales as the error levels change drastically with the length of the sample trajectory.

    Fig. \ref{fig.Eg2} illustrates that the learning accuracy and  the distribution function $\rho_T$ depend on the length of the sampling trajectory. As $T\rightarrow \infty$, the probability $\rho_T$ converges to the stationary probability $\rho$. The candidate hypothesis space for this experiment is coercive. Thus, Theorem \ref{thm.learn.coercive} can be applied, which implies that the convergence happens with respect to ${L^{2, *}_{{\rho}}}$--norm. Despite the fact that the estimation diverges from the original couplings in some regions, one can observe that $\nu_\rho(r)$ converges to zero almost everywhere on the interval $[0,\, R]$. Hence, the estimator converges over those regions where information are found in abundance.

    \begin{table}[th!]
        \centering
        \begin{tabular}{||c |c c ||} 
         \hline
         T & $\tE_T(\hat{\phi}_T)$ &  $\|\phi-\hat{\phi}_T\|_{L^{2, *}_{{\rho}}}$ \\ [1ex] 
         \hline\hline
         $10^2$ & $0.2580$ & $3.8379$  \\ 
         $10^3$ & $0.0815$ & $0.7556$  \\
         $10^4$ & $0.0259$ & $0.5574$  \\
         $10^5$ & $0.0083$& $0.3606$  \\ 
         $\infty$ & $0.0011$ & $0.3060$ \\ [.05ex] 
         \hline
        \end{tabular}
        \caption{This Table illustrates the error functional, $\tE(\hat\phi_T)$ and the distance of empirical estimator and the original function, $\|\phi-\hat{\phi}\|_{L^{2, *}_{{\rho}}}$,  
         with different trajectory lengths $T$ for the second example in Subsection \ref{sim.example.b}}
        \label{table:ex2}
    \end{table}

The numerical values of the learning error  $\tE_T(\hat{\phi}_T)$, the distance of the estimated and the original function \linebreak[4] $\|\phi-\hat{\phi}_T\|_{L^{2, *}_{{\rho}}}$ are reported in Table \ref{table:ex2}. In spite the fact that both empirical error $\tE_T$ and the distance between  the empirical estimator and the original coupling function are strictly decreasing as sample trajectory length increases, they both reach a saturation level. However, the reason why these parameters  saturate is different in each case. According to \eqref{dfn.error.infty},  $\tE_T(\psi)$  contains two parts, where the first term depends on the difference between the candidate coupling function and the original one, while the second term,  that is given by $\sigma^2/N_e$, depends solely on the noise variance. On the contrary, the source of saturation in $\|\phi-\hat{\phi}_T\|_{L^{2, *}_{{\rho_T}}}$ originates from the limitation of the hypothesis space $\Hf$ to approximate the original coupling function, which is discussed in  Theorem \ref{thm.learn.coercive}. The distance between the empirical estimator from the original coupling function tends to the distance between the original coupling function and $\Hf$ as the original coupling function does not belong to  $\Hf$. Fig. \ref{fig.ex2.Best} illustrates the best coupling function that one can learn from  $\Hf$ alongside the expected probability density $\rho$. This function is obtained by taking samples from $\phi$ with respect to the probability density function $\rho$. It can be seen that  $\hat \phi_T$ is close to the best possible approximation.

	\section{Conclusions}
	We develop a framework to learn nonlinear coupling functions for  a class of stochastic dynamical networks using only one, but long enough, sample trajectory. This requires us to prove that such networks can generate geometrically ergodic trajectories in order to ensure that collecting new sample points along the same trajectory will contain useful information for learning purposes. We obtain several error bounds for the learning accuracy and show that as the length of the sample trajectory increases, the quality of learning improves to its limit.  



\printbibliography

@article{cucker2007mathematics,
  title={On the mathematics of emergence},
  author={Cucker, Felipe and Smale, Steve},
  journal={Japanese Journal of Mathematics},
  volume={2},
  number={1},
  pages={197--227},
  year={2007},
  publisher={Springer}
}

@ARTICLE{keith-2021,
  author={ Kurt, M. and  Mivehchi, A. and  Moored, K.},
  journal={Fluids}, 
  title={High-Efficiency Can Be Achieved for Non-Uniformly Flexible Pitching Hydrofoils via Tailored Collective Interactions}, 
  year={2021},
  volume={6:233},
  number={7},
  pages={},
}

@ARTICLE{dan-potential,
  author={Rimon, E. and Koditschek, D.E.},
  journal={IEEE Transactions on Robotics and Automation}, 
  title={Exact robot navigation using artificial potential functions}, 
  year={1992},
  volume={8},
  number={5},
  pages={501-518},
  doi={10.1109/70.163777}}

@article{STROGATZ20001,
  author={Strogatz, S.H.},
title = {From Kuramoto to Crawford: exploring the onset of synchronization in populations of coupled oscillators},
journal = {Physica D: Nonlinear Phenomena},
volume = {143},
number = {1},
pages = {1-20},
year = {2000}}

@incollection{strogatz1994norbert,
  title={Norbert Wiener’s brain waves},
  author={Strogatz, Steven H},
  booktitle={Frontiers in mathematical biology},
  pages={122--138},
  year={1994},
  publisher={Springer}
}

@article{cucker2007emergent,
  title={Emergent behavior in flocks},
  author={Cucker, Felipe and Smale, Steve},
  journal={IEEE Transactions on automatic control},
  volume={52},
  number={5},
  pages={852--862},
  year={2007},
  publisher={IEEE}
}

@article{vicsek1995novel,
  title={Novel type of phase transition in a system of self-driven particles},
  author={Vicsek, Tam{\'a}s and Czir{\'o}k, Andr{\'a}s and Ben-Jacob, Eshel and Cohen, Inon and Shochet, Ofer},
  journal={Physical review letters},
  volume={75},
  number={6},
  pages={1226},
  year={1995},
  publisher={APS}
}

@book{gardiner1985handbook,
  title={Handbook of stochastic methods},
  author={Gardiner, Crispin W and others},
  volume={3},
  year={1985},
  publisher={springer Berlin}
}

@book{tong1990non,
  title={Non-linear time series: a dynamical system approach},
  author={Tong, Howell},
  year={1990},
  publisher={Oxford University Press}
}

@article{cline1999geometric,
  title={Geometric ergodicity of nonlinear time series},
  author={Cline, Daren BH and Pu, Huay-min H},
  journal={Statistica Sinica},
  pages={1103--1118},
  year={1999},
  publisher={JSTOR}
}

@article{kristensen2005geometric,
  title={Geometric ergodicity of a class of Markov chains with applications to time series models},
  author={Kristensen, Dennis},
  journal={Available at SSRN 831068},
  year={2005}
}

@book{meyn2012markov,
  title={Markov chains and stochastic stability},
  author={Meyn, Sean P and Tweedie, Richard L},
  year={2012},
  publisher={Springer Science \& Business Media}
}

@article{jensen2007law,
  title={On the law of large numbers for (geometrically) ergodic Markov chains},
  author={Jensen, S{\o}ren Tolver and Rahbek, Anders},
  journal={Econometric Theory},
  volume={23},
  number={4},
  pages={761--766},
  year={2007},
  publisher={Cambridge University Press}
}

@article{lemanczyk2021general,
  title={General Bernstein-like inequality for additive functionals of Markov chains},
  author={Lema{\'n}czyk, Micha{\l}},
  journal={Journal of Theoretical Probability},
  volume={34},
  number={3},
  pages={1426--1454},
  year={2021},
  publisher={Springer}
}

@article{cucker2008flocking,
  title={Flocking in noisy environments},
  author={Cucker, Felipe and Mordecki, Ernesto},
  journal={Journal de math{\'e}matiques pures et appliqu{\'e}es},
  volume={89},
  number={3},
  pages={278--296},
  year={2008},
  publisher={Elsevier}
}

@article{lu2019nonparametric,
  title={Nonparametric inference of interaction laws in systems of agents from trajectory data},
  author={Lu, Fei and Zhong, Ming and Tang, Sui and Maggioni, Mauro},
  journal={Proceedings of the National Academy of Sciences},
  volume={116},
  number={29},
  pages={14424--14433},
  year={2019},
  publisher={National Acad Sciences}
}

@article{bongini2017inferring,
  title={Inferring interaction rules from observations of evolutive systems I: The variational approach},
  author={Bongini, Mattia and Fornasier, Massimo and Hansen, Markus and Maggioni, Mauro},
  journal={Mathematical Models and Methods in Applied Sciences},
  volume={27},
  number={05},
  pages={909--951},
  year={2017},
  publisher={World Scientific}
}

@article{lu2021learning,
  title={Learning interaction kernels in heterogeneous systems of agents from multiple trajectories.},
  author={Lu, Fei and Maggioni, Mauro and Tang, Sui},
  journal={J. Mach. Learn. Res.},
  volume={22},
  pages={32--1},
  year={2021}
}

@article{hardt2016gradient,
  title={Gradient descent learns linear dynamical systems},
  author={Hardt, Moritz and Ma, Tengyu and Recht, Benjamin},
  journal={arXiv preprint arXiv:1609.05191},
  year={2016}
}

@article{dean2020sample,
  title={On the sample complexity of the linear quadratic regulator},
  author={Dean, Sarah and Mania, Horia and Matni, Nikolai and Recht, Benjamin and Tu, Stephen},
  journal={Foundations of Computational Mathematics},
  volume={20},
  number={4},
  pages={633--679},
  year={2020},
  publisher={Springer}
}

@article{brunton2016discovering,
  title={Discovering governing equations from data by sparse identification of nonlinear dynamical systems},
  author={Brunton, Steven L and Proctor, Joshua L and Kutz, J Nathan},
  journal={Proceedings of the national academy of sciences},
  volume={113},
  number={15},
  pages={3932--3937},
  year={2016},
  publisher={National Acad Sciences}
}

@inproceedings{foster2020learning,
  title={Learning nonlinear dynamical systems from a single trajectory},
  author={Foster, D. and Sarkar, T. and Rakhlin, A.},
  booktitle={Learning for Dynamics and Control},
  pages={851--861},
  year={2020},
  organization={PMLR}
}

@article{chen2021maximum,
  title={Maximum likelihood estimation of potential energy in interacting particle systems from single-trajectory data},
  author={Chen, Xiaohui},
  journal={Electronic Communications in Probability},
  volume={26},
  pages={1--13},
  year={2021},
  publisher={Institute of Mathematical Statistics and Bernoulli Society}
}

@article{poggio2002mathematical,
  title={On the mathematical foundations of learning},
  author={Poggio, Tomaso and Shelton, Christian R},
  journal={American Mathematical Society},
  volume={39},
  number={1},
  pages={1--49},
  year={2002},
  publisher={Citeseer}
}

@inproceedings{dean2019safely,
  title={Safely learning to control the constrained linear quadratic regulator},
  author={Dean, Sarah and Tu, Stephen and Matni, Nikolai and Recht, Benjamin},
  booktitle={2019 American Control Conference (ACC)},
  pages={5582--5588},
  year={2019},
  organization={IEEE}
}

@book{mohri2018foundations,
  title={Foundations of machine learning},
  author={Mohri, Mehryar and Rostamizadeh, Afshin and Talwalkar, Ameet},
  year={2018},
  publisher={MIT press}
}

@book{cucker2007learning,
  title={Learning theory: an approximation theory viewpoint},
  author={Cucker, Felipe and Zhou, Ding Xuan},
  volume={24},
  year={2007},
  publisher={Cambridge University Press}
}

@article{roberts2008variance,
  title={Variance bounding Markov chains},
  author={Roberts, Gareth O and Rosenthal, Jeffrey S},
  journal={The Annals of Applied Probability},
  volume={18},
  number={3},
  pages={1201--1214},
  year={2008},
  publisher={Institute of Mathematical Statistics}
}

@article{haggstrom2007variance,
  title={On variance conditions for Markov chain CLTs},
  author={Haggstrom, Olle and Rosenthal, Jeffrey},
  journal={Electronic Communications in Probability},
  volume={12},
  pages={454--464},
  year={2007},
  publisher={Institute of Mathematical Statistics and Bernoulli Society}
}

@book{christensen2003introduction,
  title={An introduction to frames and Riesz bases},
  author={Christensen, Ole and others},
  volume={7},
  year={2003},
  publisher={Springer}
}

@inproceedings{liu2021risk,
  title={Risk of Cascading Failures in Time-Delayed Vehicle Platooning},
  author={Liu, Guangyi and Somarakis, Christoforos and Motee, Nader},
  booktitle={2021 60th IEEE Conference on Decision and Control (CDC)},
  pages={4841--4846},
  year={2021},
  organization={IEEE}
}

	\section{Appendix A. Related Theorems}\label{sec:appendix}

    In this appendix we consider the Markov chain evolve by equation \eqref{eqn.Markov.chain} i.e.,
    \begin{equation}
		Y_{t+1} = G(Y_t,W_{t+1}) , \quad t\in \mathbb{Z}_+,\nonumber
	\end{equation}
    where $G:\mathcal{Y} \times \mathcal{O}_W \rightarrow \mathcal{Y}$, $ \{Y_t\}$ is a  $q$-dimensional Markov chain  taking values in $\mathcal{Y} \subset \R^q$, and noise $W_t$ is $\R^p$-valued i.i.d random variables with support $\mathcal{O}_W$. 

	For ease of notation let us define  functions $G_n, n\ge 1$ inductively by
	$$G_1(Y,W_1)=G(Y,W_1)$$
	and 
	$$G_{n+1}(Y,W_1,\dots,W_{n+1})=G(G_{n}(Y,W_1,\dots,W_{n}),W_{n+1})
	$$
     for $n\ge 1$. In the following theorem, we recall a
     sufficient condition for 
     the the Markov chain $\{Y_t\}$ to be geometrically ergodic.

	\begin{thm}  \label{thm.ergodic} For the Markov chain  $\{Y_t\}$ evolve by difference equation \eqref{eqn.Markov.transition}, we assume that the following conditions are satisfied:
		\begin{itemize}
			\item[(A0)] The sequence $W_t \in \mathcal{O}_W$ is an i.i.d random variables  on $\R^p$, and $W_t$ are independent of the initial condition, $Y_0$. 
			The marginal distribution is given by a lower semicontinuous density function $\mathfrak{g}$ w.r.t. Lebesgue measure which has support $\mathcal{O}_W=\{y\in \mathcal{Y} | \mathfrak{g}(y) >0\}$. 
			
			\item[(A1)] There exist a pair $(Y^*,W^*) \in \mathcal{Y}  \times \mathcal{O}_W $ such that 
			\[Y^* = \lim_{n \rightarrow \infty} G_n(Y,W^*,W^*,\dots,W^*),\quad Y\in \mathcal{Y} .\]
			
			\item[(A2)] The function $(Y,\,W) \mapsto G(Y,\,W)$ is continuous on $\mathcal{Y} \times \mathcal{O}_W$ and differentiable at $(Y^*,W^*)$.
			
			\item[(A3)] The matrices $\A = \partial_Y G(Y^*,W^*)$ have spectral radius $\max\{\lvert \lambda_{\A}\lvert\}<1$, and $\B = \partial_W G(Y^*,W^*)$  satisfies
			\[\text{rank}(\C_q)=\text{rank}\left[\A^{q-1}\B|~\A^{q-2}\B|~\cdots |~\A\B|~\B\right]=q.\]
			
			\item[(A4)] There exist a continuous function $V: \mathcal{Y} \rightarrow [0,\infty)$ satisfying $V(Y)\rightarrow \infty$ as $\norm{Y} \rightarrow \infty$, and constants $c_V>0,\, \rho \in (0,\,1)$, and $T \geq 1$ such that
			\begin{equation*}
				\E[V(Y_{T}\,|~Y_0=y)]\leq \rho \V(y) + C_V, \qquad \forall\, y \in \mathcal{Y}.
			\end{equation*}
		\end{itemize}
		Then $\{Y_t\}$ is geometrically ergodic chain that has a stationary solution with an invariant distribution $\pi$. Furthermore $$\E_\pi[V(Y_t)]<\infty,$$ 
		where $\E_\pi[\cdot]$ represent the expectation with respect to stationary distribution $\pi$. 
	\end{thm}
    \begin{proof}
          We refer to reference \cite{kristensen2005geometric} for  complete proof. 
    \end{proof}

 	\begin{thm} \label{thm.LLN}
    	Assume that $\{Y_t\}$ is a geometrically ergodic  chain with invariant probability $\pi$ then the law of large numbers (LLN) holds for any function g satisfying $\E_\pi[\lvert g\lvert]<\infty$ i.e. 
		\begin{equation}\label{MP.LLN}
		    \lim_{n \rightarrow \infty }\frac{1}{n} \sum_{i=1}^n g(Y_n) = \E_\pi[g].
		\end{equation}
	\end{thm}
    \begin{proof}
        We refer to references \cite{jensen2007law,meyn2012markov} for  complete proof.
    \end{proof}

    To study nonlinear dynamics on a network with the presence of bounded noises, we can need a concentration inequality for geometrically ergodic Markov chains.

    \begin{thm}\label{thm.Concentrtion}
	    Let $\{Y_t\}$ be a geometrically
		ergodic Markov chain with state space $\mathcal{Y}$,  $\pi$ be its unique stationary probability
		measure, and \linebreak[4] let $g:\mathcal{Y} \rightarrow \R$  be a bounded measurable function such that $\E_\pi [g] =0$.  Then for  $Y_0 \in \mathcal{Y}$,  one can find constants $\mathcal{C}_e,\,\tau$ depending only on $Y_0$ and the transition probability $\Prb(\cdot,\cdot)$ such that for all $\gamma>0$,
		\begin{align}\label{thm.Concentrtion.ineq}
			&\Prb_{Y_0} \left(\Big\lvert\sum_{i=0}^{n-1} g(Y_i)\Big\lvert > \gamma \right)\leq \nonumber\\
			& \hspace{1cm} \mathcal{C}_e \exp\left(-\frac{\gamma^2}{32n\sigma_\mathcal{M}^2+\tau \gamma \norm{g}_\infty\log (n)}  \right),
		\end{align}
	where 
	\begin{equation}\label{thm.Concentrtion.asymvar}
	    \sigma_{\mathcal{M}}^2 =\text{Var}_\pi\big(g(Y_0)\big) + 2 \sum_{i=1}^{\infty}\text{Cov}_\pi\big(g(Y_0),\, g(Y_i)\big),
	\end{equation}
    is the asymptotic variance of the process $\{g(Y_i)\}_{i\geq 1}$ .
	\end{thm}
	\begin{proof}
	    We refer to reference \cite{lemanczyk2021general} for complete proof
	\end{proof}
	
	\begin{lem}\label{lem.variance.bound}
	The asymptotic variance is finite and bounded if function $g$ in Theorem \ref{thm.Concentrtion} satisfies
	\[\E_\pi[g^2] \leq \infty,\]
	and the Markov chain $\{\X_t\}$ is geometrically ergodic and reversible. Moreover, there exists a nonnegative constant $c_\mathcal{M}$ such that
	\begin{equation}\label{lem.variance.upbound.ineq}
	    \Sigm^2 \leq c_\mathcal{M} \Var_\pi(h).
	\end{equation}
	\end{lem}
	\begin{proof}
	    We refer to references \cite{roberts2008variance,haggstrom2007variance} for complete proof.
	\end{proof}

   \section{Appendix B. Proofs}\label{Apendix.Proof}

\textit{Proof of Proposition \ref{lem.state.bounded}}.
    	From \eqref{dfn.noise.propos} and \eqref{eqn.network.dyn.perp.main}, we have 
        \begin{equation*}	
            \|\X_{t+1}\|\le 	\zeta \norm{\X_t} + h { \omega},
    	\end{equation*}
    	holds almost surely. Applying the above  estimate repeatedly, we obtain
    	\begin{equation} \label{ineq.lem.state.upperbound}
    		\norm{\X_{t+1}}   \leq   \zeta^{t+1} \norm{\X_0}  + h {\omega} \sum_{s=0}^{t} \zeta^{s}
    		   \le  R_0  +  \frac{h\omega}{1-\zeta},
    	\end{equation}
    	holds almost surely. The last inequality follows from Assumption \eqref{asm.eigen}  that implies $\zeta < 1$. This together with the definition of $r_t^{ij}$ completes the proof. \qed
    	
\textit{Proof of Theorem \ref{thm.network.ergodic}}.
    	 Let us consider 
        \begin{equation} \label{lem.network.ergodic.eq1}
            G(x_t, w_t):= (I -h L_{\X_{t}}) \X_{t} + h\W_{t}.
        \end{equation}
        According to the Theorem \ref{thm.ergodic} in Section \ref{sec:appendix}, for proving ergodicity of Markov chain generated by \eqref{eqn.network.dyn.perp.main}  it suffices to verify conditions (A0)-(A4). Condition (A0)  holds as $\W_{t, \perp}$  is bounded for all $t\ge 0$. The equilibrium of the network is $(\X^*,\, \W^*):=(0,0)$, i.e.,  $G(\X^*,\, \W^*)=0$. Thus, (A1) is satisfied. Assumption \ref{asm.true.kernel} combined with \eqref{lem.network.ergodic.eq1} guarantees  continuity of $G$  over $\mathcal{X} \times \mathcal{W}$ and its differentiability at $(\X^*,\, \W^*)$, which implies that (A2) holds.  
        Let us define 
        \[
        A := \frac{\partial G}{\partial \X} ( 0, 0) = M_n-h L_{ 0}, \quad 
        B := \frac{\partial G}{\partial \W} (0,  0) = h M_n,
        \]
        where  $L_{0}$ is the Laplacian matrix  \eqref{eqn.dfn.laplacian} evaluated at $0$. 
        By Assumption \ref{asm.graph},
        $L_{\bf 0}$ has a simple eigenvalue at zero with the corresponding eigenspace  $\Delta$.
        Therefore,  the restriction of the matrix to  $\Delta^{\perp}$ is full rank. Hence, the requirement (A3) holds. Finally, let us consider function $V(\X)= {\X}^T {\X}$ on $\Delta^\perp$. This function tends to $\infty$ as $\|\X\|\to \infty$. It follows that
        \begin{align}
               {\mathbb E}[V(\X_{t+1})\,|\,\X_0=\bar \X] 
            & =  
            {\mathbb E}
            \big[\X_{t}^T(I-h L_{\X_t})^T(I-h L_{\X_t})\X_{t}|\X_0=\bar \X\big]
            \nonumber\\
            &  + 2h
            {\mathbb E}\big[\W_{t}^T (I-hL_{\X_t})\X_{t}|\X_0=\bar \X \big]\nonumber\\
             &  + h^2 {\mathbb E}\big[\W_{t}^T\W_{t}|\X_0=\bar \X \big]\nonumber \\
            & \le  \zeta^2 
            {\mathbb E}
            \big[V(\X_{t})|\X_0=\bar \X \big]+   h^2  \sigma^2,\qquad \nonumber
        \end{align}
        where the last inequality follows from \ref{dfn.noise.propos} and
        \eqref{xi-value}.
        Applying the above estimate repeatedly proves that
        the condition (A4)  is satisfied.\qed

\textit{Proof of Lemma \ref{lem.upbound}}
        For every $\psi_1, \psi_2\in {\mathcal H}$, \eqref{ineq.Error.Measure} can be applied to show that
        \begin{align}
             |\tE(\psi_1)-\tE(\psi_2)|\nonumber &\le
            \left(N_e^{-1}\E_\pi \left[ \norm{\F_{\psi_1-\psi_2}(\X)}^2 \right]\right)^{1/2}\nonumber\\
             &\times \left(N_e^{-1}\E_\pi  \left[\norm{\F_{\psi_1+\psi_2-2\phi}(\X)}^2 \right]\right)^{1/2}\nonumber\\
             & \le   K^2  \|\psi_1-\psi_2\|_{L^{2, *}_{\rho}} ~
            \|\psi_1+\psi_2-2\phi\|_{L^{2, *}_{\rho}}\nonumber\\
            & \le   2 K^2R \big(S_{{\mathcal H}, R}+\|\phi\|_{\infty }\big)  \|\psi_1-\psi_2\|_{L^{2, *}_{\rho}} \nonumber
        \end{align}
        holds almost surely, where the last inequality follows from \eqref{ineq.l2.linfty.bounded} and \eqref{def.SHR}. One can prove inequality \eqref{lem.upbnd.T} similarly.\qed

    \textit{Proof of Theorem \ref{thm.compact}}.
            By definition \eqref{dfn.g.psi}, it is evident that $\E_\pi[g_\psi] =0 $.  From \eqref{ineq.diff.infty}, it follows that 
            $\norm{g_\psi}_\infty\leq M$, where  $M=8K^2R^2S^2$.
            Since $\X_t$ is geometrically ergodic,  we can apply concentration inequality \eqref{thm.Concentrtion.ineq} in  Theorem \ref{thm.Concentrtion} in the Appendix section to obtain an upper bound for probability of $| L_T(\psi) | $ being greater than $\epsilon > 0$, i.e.,
            \begin{equation}\label{thm.compact.prf.concentration}
             \hspace{-0.25cm}   \Prb\left\{ \abs{L_T(\psi)}\geq \epsilon\right\}\leq
                C \exp\left\{\frac{-\epsilon^2T}{32\Sigm^2(\psi)+\tau\epsilon M \log(T)}\right\}
            \end{equation}
            where $C_e$ and $\tau$ are constants with respect to parameters of the network \eqref{eqn.network.dyn.perp.main}. 
            Lemma \ref{lem.variance.bound} asserts that the asymptotic variance $\Sigm^2(\psi)$ is bounded for every $\psi \in \Hf$. If we define $\sigma_\Hf := \sup_{\psi \in \Hf} \Sigm(\psi)$, then
            \begin{equation*}
             \frac{-\epsilon^2T}{32\Sigm^2(\psi)+\tau\epsilon M \log(T)} \leq \frac{-\epsilon^2T}{32\sigma_\Hf^2+\tau\epsilon M \log(T)}. 
            \end{equation*}
      Suppose that $l =\mathcal{N}(\Hf,\frac{\epsilon S}{2M})$ is the minimum number of disks $\mathcal{D}_1, \ldots, \mathcal{D}_l $ to cover $\Hf$, where \[\mathcal{D}_j=\left\{ \psi \in \Hf \hspace{0.03cm} \left| \hspace{0.03cm} \|\psi - \psi_j \|_{\infty} \leq \epsilon S/2M \right.\right\}.\] 
For every $\psi \in \Dj$, the result of Lemma \ref{lem.upbound} implies that 
            \[
            \abs{L_T(\psi)-L_T(\psi_j)} \leq \frac{M}{2S}\norm{\psi-\psi_j}_\infty \leq \frac{\epsilon}{4}.
            \]
            Then, we get
            \begin{equation} 
                \sup_{\psi \in \Dj}\abs{ L_T(\psi)}\geq \frac{\epsilon}{2} \Rightarrow  \abs{L_T(\psi_j)}\geq \frac{\epsilon}{4}\nonumber
            \end{equation}
            for all $1 \leq j \leq l$ and $1\leq T$, which translates to 
            \begin{equation}\label{thm.compact.Prf.ineq2}
                \Prb\left\{\sup_{\psi \in \Dj}\abs{ L_T(\psi)}\geq \frac{\epsilon}{2}\right\} \leq \Prb\left\{\abs{L_T(\psi_j)}\geq \frac{\epsilon}{4}\right\}.
            \end{equation}
            
             Since $\{\Dj\}_{j=1}^l$ is a covering of the hypothesis space $\Hf$, we have$\Hf \subset \cup_{j} \Dj$, which results in 
             \begin{equation}\label{thm.compact.Prf.ineq3}
                 \Prb\left\{\sup_{\psi \in \Hf}\abs{ L_T(\psi)}\geq \epsilon\right\} \leq \sum_{j=1}^l \Prb\left\{\sup_{\psi \in \Dj}\abs{ L_T(\psi)}\geq \epsilon\right\}.
             \end{equation}
             The inequality \eqref{thm.compact.Prf.ineq3} extends  the local probability approximation on each $\Dj$ to the hypothesis space $\Hf$. For some $\delta \in [0,1]$, let us assume that
             \begin{equation*}
                 \Prb \left\{\sup_{\psi \in \Hf}\abs{ L_T(\psi)}\geq \epsilon/2 \right\} \geq 1-\delta.
             \end{equation*}
            Since $\hat{\phi}_T,\hat{\phi} \in \Hf$, inequalities
             \begin{equation*}
                 \tE(\hat{\phi}_T)\leq \tE_T(\hat{\phi}_T) +\frac{\epsilon}{2}~~~\textrm{and}~~~ \tE_T(\hat{\phi})\leq \tE(\hat{\phi}) +\frac{\epsilon}{2}
             \end{equation*}
            hold with probability at least $1-\delta$. It follows that
            \begin{equation*}
                \tE(\hat{\phi}_T) \leq \tE_T(\hat{\phi}_T)+\frac{\epsilon}{2} \leq \tE_T(\hat \phi) +\frac{\epsilon}{2} \leq \tE(\hat \phi) +\epsilon
            \end{equation*}
            holds with the same probability using the fact that $\tE_T(\hat \phi_T)\leq \tE_T(\hat \phi)$. Finally, we conclude that
            \begin{equation*}
                \Prb\Big\{\tE(\hat{\phi}_T) -\tE(\hat \phi) \leq  \epsilon\Big\}\leq 1-\delta,
            \end{equation*}
            Therefore, replacing $\frac{\epsilon}{4}$ by $\epsilon$ in  \eqref{thm.compact.prf.concentration}  completes the proof.\qed
            Before providing a proof for Theorem \ref{thm.learn.coercive}, we establish some essential results.  For every $\psi \in \Hf$, we define 
	\begin{equation}\label{def.DT}
D_T(\psi):= \tE_{T}(\psi)-\tE_{T}(\hat{\phi})~~\textrm{and}~~
	    D(\psi):= \tE(\psi)-\tE(\hat{\phi}).
	\end{equation}

	
	\begin{lem}\label{lem.consentration}
For every $\alpha \in (0,\,1)$, $\epsilon > 0$, and $\psi \in \Hf$,  we have 
        \begin{align}\label{lem.consentration.Prb}
       \hspace{-0.2cm}     \Prb\left\{\frac{D(\psi)-D_T(\psi)}{D(\psi)+\epsilon}\geq \alpha\right\} &\leq\\
            &\hspace{-2cm} C \exp\left\{\frac{-T\alpha^2\epsilon}{16K^2R^2S^2\left(\frac{128K^2 c_\mathcal{M}}{c_\Hf} + \tau   \log(T) \right)}\right\}, \nonumber
        \end{align}
        where $C,\,\tau,c_\mathcal{M}$  are constants with respect to the rate of ergodidcty of \eqref{eqn.network.dyn.perp.main}.
	\end{lem}
    \begin{proof}
         For every $\psi  \in \Hf$, let us define $\bgp:\R^{dn} \rightarrow \R$ by
         \begin{equation*}
             \bgp(\X_t):= \tE_{\x_t}(\psi)-\tE_{\x_t}(\hat \phi)-\tE(\psi)+\tE(\hat\phi).
         \end{equation*}
         It can be verified that $\E_\pi[\bgp] =0$ and $\norm{\bgp}_\infty \leq M$, where\linebreak[4] $M=8K^2R^2S^2$. 
         As random variable $\x_t$ is geometrically ergodic, we can apply \eqref{thm.Concentrtion.ineq} to obtain an upper bound for the confidence level. Let us denote the asymptotic variance of $\bar{g}(\X_t)$ by $\bar \sigma_\mathcal{M}(\psi)$. Thus,
         using Lemma \ref{lem.variance.bound},  the associated asymptotic variance is bounded by $c_\mathcal{M}\Var_\pi(\bgp)$. As a result, we have
         \begin{align*}
             \bar \sigma_\mathcal{M}^2(\psi)&\leq c_\mathcal{M} \Var_\pi(\bgp) \leq c_\mathcal{M}\E_\pi \left[\big(\tE_{\X_t}(\psi)-\tE_{\X_t}(\hat \phi)\big)2\right]\\
             &\leq c_\mathcal{M} K^4\|\psi-\hat\phi\|_{L^{2, *}_{\rho}}^2 \|\psi+\hat\phi-2\phi\|_{L^{2, *}_{\rho}}\\
             &\leq 64c_\mathcal{M}S^2 R^2 K^4\|\psi-\hat\phi\|_{L^{2, *}_{\rho}}^2\\
             &\leq \frac{8c_\mathcal{M}MK^2}{c_\Hf} (\tE(\psi)-\tE(\hat \phi)) = \frac{8c_\mathcal{M}MK^2}{c_\Hf}D(\psi),
         \end{align*}
         where the last inequality comes from the coercivity condition \eqref{dfn.coercivity.condition-1}.
         Applying Theorem \ref{thm.Concentrtion} gives us  
         \begin{align}\label{lem.consentration.prf.ineq}
             \Prb\left\{\frac{D(\psi)-D_T(\psi)}{D(\psi)+\epsilon}\geq \alpha\right \}  &\leq\\
                &\hspace{-2.5cm} C \exp\left\{\frac{-\alpha^2T^2(D(\psi)+\epsilon)^2}{32T\bar \sigma_\mathcal{M}^2(\psi)  + 2\tau \alpha T(D(\psi)+\epsilon) \log(T) }\right\}. \nonumber
         \end{align}
         A closer investigation of the exponential rate in \eqref{lem.consentration.prf.ineq} reveals that 
         \begin{align*}
             &\frac{\alpha^2T^2(D(\psi)+\epsilon)^2}{32T\bar \sigma_\mathcal{M}^2(\psi)  + 2\tau \alpha TM(D(\psi)+\epsilon) \log(T) }  \\ 
             &\geq \frac{T\alpha^2(D(\psi)+\epsilon)^2}{32\left(\frac{8c_\mathcal{M}MK^2}{c_\Hf}D(\psi)\right)  + 2\tau \alpha  M (D(\psi)+\epsilon) \log(T) }\\
             &\geq \frac{T\alpha^2(D(\psi)+\epsilon)}{\frac{256  c_\mathcal{M}MK^2}{c_\Hf}  + 2\tau \alpha  M  \log(T) }\\
             &\geq \frac{T\alpha^2\epsilon}{\frac{256  c_\mathcal{M}MK^2}{c_\Hf}  + 2\tau   M  \log(T) }.
         \end{align*}
The last inequality follows from  the fact that $D(\psi)>0$ for all $\psi \in \Hf$ and $0<\alpha<1$. The above inequality together with the concentration inequality \eqref{lem.consentration.prf.ineq} completes the proof.
    \end{proof}
    The following Proposition extend the result of Lemma \ref{lem.consentration} to the entire hypothesis space $\Hf$.

    \begin{Prop}\label{Prop.extension}
        For all $\epsilon > 0$ and $\alpha \in (0,1)$, we have
        \begin{align}\label{Prop.extension.ineq}
          \hspace{-0.5cm}  \Prb \left\{\sup_{\psi \in \Hf} \frac{D(\psi)-D_T(\psi)}{D(\psi)+\epsilon}\geq 3\alpha\right\} \leq C \mathcal{N}\left(\Hf,\frac{\alpha \epsilon}{4 K^2 R^2 S}\right)  & \nonumber\\
            &\hspace{-7.5cm} 
            \times \exp\left\{ \frac{-T\alpha^2\epsilon}{16K^2S^2R^2\left(\frac{128  c_\mathcal{M}K^2}{c_\Hf}  + \tau    \log(T)\right) }\right\}, 
        \end{align}
        where constants $c_\mathcal{M},\, \tau,\, C$ are as in Theorem \ref{thm.learn.coercive}.
    \end{Prop}
     \begin{proof}
        According to Lemma \ref{lem.consentration}, inequality 
        \begin{equation} \label{Prop.extension.prf1}
             \frac{D(\psi_1)-D_T(\psi_1)}{D(\psi_1)+\epsilon} > \alpha
        \end{equation}
        holds with probability at most 
        \begin{equation*}
            \delta^* = C \exp\left\{\frac{-T\alpha^2\epsilon}{16K^2R^2S^2 \left(\frac{128K^2 c_\mathcal{M}}{c_\Hf} + \tau   \log(T) \right)}\right\}.
        \end{equation*}

If $\psi_1,\psi_2 \in \Hf$ such that  $\|\psi_1 - \psi_2 \|_\infty \leq \frac{\alpha\epsilon}{4K^2R^2S}$, then 
         \begin{equation*}
             \frac{D(\psi_2)-D_T(\psi_2)}{D(\psi_2)+\epsilon} = \frac{L_T(\psi_2)-L_T(\psi_1)}{D(\psi_2)+\epsilon} +\frac{L_T(\psi_1)-L_T(\hat \phi)}{D(\psi_2)+\epsilon}.
         \end{equation*}
Using  \eqref{ineq.diff.infty}, the first term  is bounded by 
        \begin{align*}
             \frac{L_T(\psi_2)-L_T(\psi_1)}{D(\psi_2)+\epsilon} & \leq \frac{4K^2R^2S \norm{\psi_1-\psi_2}_\infty}{D(\psi_2)+\epsilon} \\
             & \leq \frac{\alpha \epsilon}{D(\psi_2)+\epsilon}\leq\alpha
        \end{align*}
 with probability at most $\delta^*$, where the last inequality is a consequence of $D(\psi_2)\geq0$. For the second term, by applying Lemma \ref{lem.upbound} and utilizing the fact that $0 < \alpha <1 $, we have
        \begin{equation*}
            \tE(\psi_1)-\tE(\psi_2)\leq 2K^2R^2S^2\norm{\psi_1-\psi_2}_\infty \leq \frac{\alpha\epsilon}{2} \leq \epsilon,
        \end{equation*}
        which implies that $\frac{D(\psi_1)+\epsilon}{D(\psi_2)+\epsilon} \leq 2$. Therefore, 
        \begin{equation*}
            \frac{L_T(\psi_1)-L_T(\hat \psi)}{D(\psi_2)+\epsilon}=\frac{D(\psi_1)-D_T(\psi_1)}{D(\psi_2)+\epsilon} \leq    \alpha \frac{D(\psi_1)+\epsilon}{D(\psi_2)+\epsilon} \leq 2\alpha.
        \end{equation*}
The combination of the above inequalities implies that if \eqref{Prop.extension.prf1} holds with probability at most $\delta^*$, then 
         \begin{equation} \label{Prop.extension.prf2}
             \frac{D(\psi_2)-D_T(\psi_2)}{D(\psi_2)+\epsilon} > 3\alpha
        \end{equation}
    holds with the same probability. One can complete the proof similar to the proof of Theorem \ref{thm.compact} by employing the covering disks and extending the local results on each disk to the entire $\Hf$. 
     \end{proof}
    
    
   \textbf{Proof of Theorem \ref{thm.learn.coercive}.}
By utilizing Proposition \ref{Prop.extension} with $\alpha = \frac{1}{6}$, it follows that  
\begin{equation}\label{thm.learn.coercive.prf1}
      \frac{D_\infty(\psi)-D_T(\psi)}{D_\infty(\psi)+\epsilon} < \frac{1}{2}
        \end{equation}
for all large enough  $T$, holds for all $\psi \in \Hf$ with probability at least 
       $1-\delta$.
      Then, it suffices to prove that
      \begin{equation}\label{thm.learn.coercive.prf2}	       \|\hat{\phi}_T-{\phi}\|^2_{L^{2, *}_{\rho}} 
	       \leq  \frac{2\epsilon }{c_\Hf}+ \left(4+\frac{4K}{c_\Hf}\right) \inf_{\psi \in \Hf}   \big\|\phi-\psi\|^2_{L^{2, *}_{\rho}}.
	   \end{equation}
If $\psi = \hat{\phi}_T$, then 
        \[
            D(\hat{\phi}) < 2 D_T(\hat{\phi}_T)+\epsilon.
        \]
From the definition of  $D(\cdot)$ and $D_T(\cdot)$ in \eqref{def.DT}, we have 
         \[
            \tE(\hat{\phi}_T) - \tE(\hat{\phi})  < 2 (\tE_T(\hat{\phi}_T) - \tE_T(\hat{\phi}))+\epsilon \leq \epsilon,
        \]
       where  the last inequality holds based on the fact that $\hat{\phi}_T$  minimizes $\tE_T(\cdot)$, which implies  $\tE_T(\hat{\phi}_T) - \tE_T(\hat{\phi}) \leq 0$. From \eqref{rem.coercivity.eq2}, one has
	    \[
	    c_\Hf \big\|\hat{\phi}_T-\hat{\phi}\big\|^2_{L^{2, *}_{\rho}} \leq \tE(\hat{\phi}_T) - \tE(\hat{\phi}) < \epsilon.
	    \]
        On the other hand, for any $\psi \in \Hf$ we have
        \begin{eqnarray*}
         & & \hspace{-1.2cm} \big\|\hat{\phi}-{\phi}\big\|^2_{L^{2, *}_{\rho}} \leq  2 \|\psi-{\phi}\big\|^2_{L^{2, *}_{\rho}}+2\big\|\hat{\phi}-\psi\|^2_{L^{2, *}_{\rho}}\\ 
           &\leq & 2\big\|{\phi}-\psi\|^2_{L^{2, *}_{\rho}}+\frac{2}{c_\Hf} \big(\tE({\psi}) - \tE({\phi}) + \tE({\phi}) - \tE(\hat{\phi})\big) \\
           &\leq & \left(2+\frac{2K}{c_\Hf}\right)  \big\|{\phi}-\psi\|^2_{L^{2, *}_{\rho}},
        \end{eqnarray*}
        where the second inequality  holds by coercivity condition and the last inequality follows from the fact that $\tE({\phi}) \le  \tE(\hat{\phi}_\infty)$ and
        the argument used to prove \eqref{rem.coercivity.eq2}.
        Taking infinium with respect to $\psi \in \Hf$ results in
        \begin{equation}
          \big\|\hat{\phi}-{\phi}\big\|^2_{L^{2, *}_{\rho}} \leq \left(2+\frac{2K}{c_\Hf}\right) \inf_{\psi \in \Hf}   \big\|\phi-\psi\|^2_{L^{2, *}_{\rho}}.
        \end{equation}
	    Finally, for the distance between $\hat{\phi}_T$ and  $\phi$, inequality
	   \begin{eqnarray*}
	       \|\hat{\phi}_T-{\phi}\|^2_{L^{2, *}_{\rho}} & \leq & 2\|{\phi}-\hat{\phi}\|^2_{L^{2, *}_{\rho}}
	       +2\|\hat{\phi}_T-\hat{\phi}\|^2_{L^{2, *}_{\rho}} \\
	       & \leq & \frac{2\epsilon }{c_\Hf}+\Big (4+\frac{4K}{c_\Hf}\Big) \inf_{\psi \in \Hf}   \big\|\phi-\psi\|^2_{L^{2, *}_{\rho}},
	   \end{eqnarray*}
	   holds with probability at leas $1-\delta$,
	   which proves \eqref{them.learn.coercive.Prb}.  \qed  
	   
    \textit{Proof of Theoerm \ref{thm.convergence.rate}}
       From Theorem \ref{thm.learn.coercive} and the fact that $\phi \in \mathcal{K}_{R,S}$, we have
       \begin{align*}
           \Prb\left\{\|\hat{\phi}_T - {\phi}\|_{L^{2, *}_{\rho}} \geq \epsilon \right\}\leq C \ \mathcal{N}\left(\Hf,\frac{ \epsilon^2 S c_\Hf}{3M}\right) &\times \\
            & \hspace{-4cm}\exp\left\{ \frac{-T\epsilon^2c_\Hf^2}{12M({128  c_\mathcal{M}K^2}  + \tau c_\Hf   \log(T)) }\right\}, \nonumber
       \end{align*}
       where $M=8K^2R^2S^2$. For set $\mathcal{K}_{R,S}$, we can approximate $\mathcal{N}(\mathcal{K}_{R,S},l)$ by 
       \begin{equation}\label{thm.convergence.rate.prf.ineq1}
           \mathcal{N}(\mathcal{K}_{R,S},l) \leq e^{\frac{\kappa}{l} },
       \end{equation}
       where $\kappa$  is a constant dependent on $\mathcal{K}_{R,S}$ and $K$ \cite{cucker2007learning}. It follows that
       \begin{align*}
           \Prb\{\|\hat{\phi}_T - {\phi}\|_{L^{2, *}_{\rho}} \geq \epsilon\}\leq &\\ 
            & \hspace{-3cm}\exp\left\{\frac{6M\kappa}{\epsilon^2c_\Hf S}- \frac{T\epsilon^2c_\Hf^2}{12M({128  c_\mathcal{M}K^2}  + \tau c_\Hf   \log(T)) } + c_0\right\}, \nonumber
       \end{align*}
       where $c_0 = \log(\Cm_0)$. For simplicity of our notations, let us set  $c_1=\frac{6M \kappa}{c_\Hf S}$, $c_2=\frac{Tc_\Hf^2}{12M({128  c_\mathcal{M}K^2} + \tau c_\Hf   \log(T))}$, which results in
       \begin{align*}
           \Prb\left\{\|\hat{\phi}_T - {\phi}\|_{L^{2, *}_{\rho}} \geq \epsilon \right\}\leq 
           \exp\left\{\frac{c_1}{\epsilon^2}- {c_2\epsilon^2}+ c_0\right\}. \nonumber
       \end{align*}
       Solving  $\frac{c_1}{\epsilon^2}- {c_2\epsilon^2}+ c_0 =0$ gives us $\epsilon^* =  \sqrt{\frac{c_0 +\sqrt{c_0^2+4c_1c_2
       }}{2c_2}}$. Thus, we have
       \begin{equation*}
        \Prb\left\{\|\hat{\phi}_T - {\phi}\|_{L^{2, *}_{\rho}} \geq \epsilon\right\}\leq 
           \begin{cases}
           \exp\left\{\frac{c_1}{\epsilon^2}- {c_2\epsilon^2}+ c_0\right\}& \textrm{if}~ \epsilon>\epsilon^*\\
           1& \textrm{if}~ \epsilon\leq\epsilon^*
           \end{cases}.
       \end{equation*}
Taking integral  results in
    \begin{equation}
     \E_\pi\left[\|\hat{\phi}_T - {\phi}\|_{L^{2, *}_{\rho}}\right] \leq  \gamma \sqrt[4]{\frac{128  c_\mathcal{M}K^2 + \tau c_\Hf   \log(T)}{T c_\Hf^2}},
    \end{equation}
    where $\gamma$ is a constant that depends on $S, K, R, C$. \qed

\end{document}